\documentclass[11pt,superscriptaddress,aps,amssymb,amsfonts,letterpaper]{revtex4}

\usepackage{amsmath}
\usepackage{amsthm}

\usepackage{graphicx,color,graphics}

\usepackage{algorithmic}

\usepackage[matrix,frame,arrow]{xy}
\usepackage{amsmath}
\newcommand{\bra}[1]{\left\langle{#1}\right\vert}
\newcommand{\ket}[1]{\left\vert{#1}\right\rangle}
\newcommand{\qw}[1][-1]{\ar @{-} [0,#1]}
\newcommand{\qwx}[1][-1]{\ar @{-} [#1,0]}
\newcommand{\cw}[1][-1]{\ar @{=} [0,#1]}

\newcommand{\gate}[1]{*{\xy *+<.6em>{#1};p\save+LU;+RU **\dir{-}\restore\save+RU;+RD **\dir{-}\restore\save+RD;+LD **\dir{-}\restore\POS+LD;+LU **\dir{-}\endxy} \qw}
\newcommand{\meter}{\gate{\xy *!<0em,1.1em>h\cir<1.1em>{ur_dr},!U-<0em,.4em>;p+<.5em,.9em> **h\dir{-} \POS <-.6em,.4em> *{},<.6em,-.4em> *{} \endxy}}

\newcommand{\control}{*-=-{\bullet}}

\newcommand{\ctrl}[1]{\control \qwx[#1] \qw}
\newcommand{\lstick}[1]{*!R!<.5em,0em>=<0em>{#1}}

\newcommand{\Qcircuit}{\xymatrix @*=<0em>}

\def\one{{\mathchoice {\rm 1\mskip-4mu l} {\rm 1\mskip-4mu l} {\rm
1\mskip-4.5mu l} {\rm 1\mskip-5mu l}}}
\renewcommand{\ket}[1]{|{#1}\rangle}
\renewcommand{\bra}[1]{\langle{#1}|}
\newcommand{\braket}[2]{\langle{#1}|{#2}\rangle}
\newcommand{\kets}[2]{|{#1}\rangle_{{}_{\!\!{#2}}}}

\renewcommand{\tensor}{\otimes}

\newcommand{\ketbra}[2]{ \ket{#1}_{\! #2}^{#2} \!\! \bra{#1}}
\newcommand{\ketbras}[3]{ \ket{#1}_{\! #3}^{#3} \!\! \bra{#2}}

\providecommand{\ignore}[1]{}

\newcommand{\strutlike}[1]{\phantom{#1}}

\def\cO{{\cal O}}

\def\cL{{\cal L}}

\def\({\left(}
\def\){\right)}

\newtheorem{thm}{Theorem}[section]
\newtheorem{cor}[thm]{Corollary}
\newtheorem{lem}[thm]{Lemma}
\newtheorem{dfn}[thm]{Definition}

\begin{document}

\title{Fast quantum algorithms for traversing paths of eigenstates}

\author{S. Boixo}
\email{boixo@caltech.edu}
\affiliation{California Institute of Technology, Pasadena, CA 91125, USA}

\author{E. Knill}
\email{knill@boulder.nist.gov}
\affiliation{National Institute of
Standards and Technology, Boulder, CO  80305,
USA}

 \author{R.D.  Somma}
 \email{somma@lanl.gov}
\affiliation{Los Alamos National Laboratory, Los Alamos, NM 87545, USA}

\date{\today}

\begin{abstract}
Consider a path of non-degenerate eigenstates $\ket{\psi_s}$, $0\leq
s\leq 1$, of unitary operators $U_s$ or Hamiltonians $H_s$ with minimum
eigenvalue gap $\Delta$. The eigenpath traversal problem is to
transform one or more copies of $\ket{\psi_0}$ into $\ket{\psi_1}$.
Solutions to this problem have applications ranging from quantum
physics simulation to optimization. For Hamiltonians, the conventional
way of doing this is by applying the adiabatic theorem. We give
``digital'' methods for performing the transformation that require no
assumption on path continuity or differentiability other than the
absence of large jumps. Given sufficient information about eigenvalues
and overlaps between states on the path, the transformation can be
accomplished with complexity $\cO((L/\Delta) \log(L/\epsilon))$, where
$L$ is the angular length of the path and $\epsilon$ is a specified
bound on the error of the output state. We show that the required
information can be obtained in a first set of transformations, whose
complexity per state transformed has an additional factor that depends
logarithmically on a maximum angular velocity along the path. This
velocity is averaged over constant angular distances and does not
require continuity. Our methods have
substantially better behavior than conventional adiabatic algorithms,
with fewer conditions on the path. They also improve on the previously
best digital methods and demonstrate that path length and the gap are the primary
parameters that determine the complexity of state transformation along a
path.
\end{abstract}

\maketitle

\section{Introduction}
\label{intro}

Most quantum algorithms exhibiting speedups are based on one of a
small number of basic tools for quantum problem solving. These tools
include phase estimation, which underlies quantum factoring, and
amplitude amplification, which can be used to solve search problems.
Another such tool is adiabatic state transformation (AST). Before
finding applications in quantum algorithms, AST was used in classical
algorithms for quantum physics
simulation~\cite{amara:qc1993a,finnila:qc1994a,kadowaki:qc1998a}. It
is now a key method for accessing low-energy states in proposed
quantum simulations of physics, both in ``digital'' quantum algorithms
based on the circuit model~\cite{ortiz:qc2001a,aspuru-guzik:qc2005a}
and in ``analog'' simulation techniques involving direct realization
of Hamiltonians in systems such as optical lattices
(see Ref.~\cite{lewenstein:qc2007a}). The potential of AST was recognized
in quantum computer science when it was proposed as a powerful
heuristic for solving satisfiability
problems~\cite{farhi_quantum_2000,farhi:qc2001a}. This led to the
idea of adiabatic quantum computing (AQC), which involves encoding the
output of any quantum algorithm in an adiabatically accessible ground
state~\cite{farhi_quantum_2000,aharonov_adiabatic_2003}. More
recently, it has been shown that AST can be used for quantum speedups
of Monte-Carlo
algorithms~\cite{somma_quantum_2008,wocjan_speedup_2008}.

We consider AST problems that involve transforming $\ket{\psi_0}$ into
$\ket{\psi_1}$, where these states are the endpoints of a path of
states $\ket{\psi_s}$ for $0\leq s\leq 1$. The $\ket{\psi_s}$ are
eigenstates of operators $O_s$, which may be unitary or Hermitian. For
simplicity, we focus on the unitary case $O_s=U_s$. If $O_s$ is
Hermitian, we define $U_s = e^{-iO_s}$. The eigenphase of
$\ket{\psi_s}$ with respect to $U_s$ is $\varphi_s$ and is assumed to
be non-degenerate. The gap to the nearest other eigenphase is
denoted by $\Delta_s \ge \Delta$. We make no continuity assumptions on
$\ket{\psi_s}$ or $U_s$. Thus, our formulation of the AST problem can
accommodate cases where the path is discrete, parameterized by integers
$j=0,\ldots,n$. It suffices to define $\ket{\psi_s}=\ket{\psi_j}$ for
$j/n\leq s<(j+1)/n$.

The complexity of the AST problem depends on the available
capabilities and what we know about the path. We assume that we can
prepare copies of $\ket{\psi_0}$, and that we can apply
quantum-controlled instances of $U_s$, both at unit cost. Additional
information such as lower bounds on the gaps, eigenphase ranges and
overlaps between the $\ket{\psi_s}$ may be available. An important
consequence of our results is that the main parameter that determines
the complexity of an AST problem is the angular length of the path,
which is defined as
\begin{equation}
L = \sup\left\{
      \sum_{j=1}^{n}\arccos(|\braket{\psi_{s_j}}{\psi_{s_{j-1}}}|)
      \;\Big|\;0=s_0<\ldots<s_n=1\right\}\;.
\end{equation}
If $\ket{\psi_s}$ is differentiable in $s$, then $L = \int_{0}^1
\| (\openone - \ket {\psi_s}\bra{\psi_s})\ket{\partial_s  \psi_s} \| ds$. For the purpose of making complexity
statements, let $\bar L = \max(\pi/2,L)$. Given sufficient information
about the overlaps between nearby states on the path, we show that the
complexity of an AST problem is $\cO((\bar L/\Delta)\log(\bar
L/\epsilon))$, where $\epsilon$ is a specified bound on the error with
which $\ket{\psi_1}$ is prepared. (Our complexity statements hide
constants that apply uniformly for all $L>0$, $0<\Delta\leq\pi$ and
$0<\epsilon < 1$.) The dependence on $L$ and $\Delta$ is within a
factor of at most $\log(\bar L)$ of optimal. That is, there are classes of
AST problems for which every algorithm uses $\Omega(\bar L/\Delta)$
applications of the $U_s$~\cite{boixo:qc2009b}. With less information,
we provide algorithms whose complexities have an additional factor that
depends on a maximum locally averaged angular velocity
\begin{equation}
v_{{\rm max}} =
  \sup\left\{L(s_1,s_2)/(s_1-s_2) \;\Big|\;
               0\leq s_1<s_2\leq 1,\;
               L(s_1,s_2) \ge\theta\right\}\;,
\end{equation}
where $L(s_1,s_2)$ is the angular length of the path restricted to
$[s_1,s_2]$, and $\theta$ is a constant. In order for $v_{{\rm max}}$
to be finite, the path must not have jumps of angular distance
$\theta$ or more. The average velocity over the whole path is $v_{{\rm
    avg}} = L/1$. If the set in the definition of $v_{{\rm max}}$ is
empty, set $v_{{\rm max}}=v_{{\rm avg}}$. We find that, in general,
the transformation can be accomplished with complexity
$\cO((\cL/\Delta)\log(\cL/\epsilon)) $ per copy of $\ket{\psi_1}$
produced, where $\cL = \bar L( \log(v_{{\rm max}}/v_{{\rm avg}})+2)$.
Implicit in this bound is an extra factor of $\log(\log( v_{{\rm
    max}}/v_{{\rm avg}})+2)$. It is required only when it is necessary
to transform multiple copies at the same time, which is the case when
sufficiently small ranges for the eigenphases are not yet known. The
error bound $\epsilon$ then applies to the state of all copies
together, not each copy separately.  All our complexities can be
refined if the gap varies over the path in a known way, allowing
faster traversals of parts of the path where the gap is large. We
provide general tools to analyze this situation.

Many problems require the production of a large number of copies of
$\ket{\psi_1}$, for example to obtain good precision on expectations
of observables or values of correlation functions. In these cases, the
first set of transformations can be used to get the overlap
information needed to optimize future transformations. Thus, except
for a first set of transformations, the complexity does not depend on
$v_{{\rm max}}$.

We can compare the complexities obtained here to those of other
techniques for solving AST problems. In the case where the operators
$O_s$ are Hamiltonians, $O_s=H(s)$, the best known technique is based
on the adiabatic approximation and involves evolving under $H(s)$,
changing $s$ slowly enough to ensure the desired transformation. This
technique works well in analog approaches to solving physics
simulation problems. Analyses of the adiabatic
approximation~\cite{avron_adiabatic_1987,childs_quantum_2002,hagedorn_elementary_2002,aharonov_adiabatic_2003,ambainis_elementary_2004,jansen_bounds_2007,lidar_adiabatic_2008,amin_consistency_2009,oreshkov_adiabatic_2010} show that
the total evolution time required to obtain $\ket{\psi_1}$ with error
bounded by $\epsilon$ satisfies $\tau \in \cO\left(\sup_s\left(\|
\partial_{s} H(s) \|^2 / \Delta^3 + \| \partial^2_{s} H(s) \| /
\Delta^2\right)/\epsilon\right)$, where $\| \cdot \|$ is the operator norm.
This assumes sufficient differentiability of $H(s)$. The dependence on
$\epsilon$ is much better if $H(s)$ is highly differentiable and turned
on/off slowly at the beginning and end of the path. The path length
satisfies $L \leq \sup_s\|\partial_s H(s)\|/\Delta$. Examples can be
constructed where $L=\Omega(\sup_s\|\partial_s H(s)\|/\Delta)$, so the
bound on $\tau$ has a term that may approach $L^2/\Delta$. Besides
requiring differentiability, this complexity has at least an extra
factor of $L$. In the absence of differentiability, it is possible to
randomize the evolution under $H(s)$ for an average complexity of
$\tau=\cO(\bar
L^2/(\Delta\epsilon))$~\cite{somma_quantum_2008,boixo:qc2009a}
given a discretization of the path with sufficiently uniform angular
step sizes. This technique can also be applied in an analog setting.
In general, the factor of $1/\epsilon$ in the complexities can be
improved to $\log(1/\epsilon)$ if we can verify $\ket{\psi_1}$ by
checking its eigenvalue with respect to $H(1)$.

The randomized methods in Ref.~\cite{boixo:qc2009a} are also
applicable to paths of unitaries. The paths must have a discretization
with asymptotically small angular distances between adjacent states,
implying a continuous underlying path. A method with better
complexity is in Ref.~\cite{wocjan_speedup_2008}. It is based on
Grover's fixed point search and if applied to our formulation of AST
has a complexity of $\cO(\bar L\log(\bar L/\epsilon)^2/\Delta)$, given
an appropriate discretization of the path. This method was not
analyzed for arbitrary paths, but a discretization with sufficiently
large overlaps between successive states on the path works, so
continuity is not required. Our work improves on this complexity, and
perhaps more importantly, shows that good complexity can be obtained
even when overlaps between states on the path and eigenphase ranges
are unknown. It suffices to have lower bounds on the gaps, and in the
least informed case, be assured that a certain eigenphase dominance
condition (to be defined below) applies.  \ignore{
As I understand Ref.~\cite{wocjan_speedup_2008}, there are
$\cO(\bar L)$ steps, each of which requires $\cO(\log(\bar
L/\epsilon))$ rotations about the states for a final error bounded by
$\epsilon$. Each rotation requires $\cO(\log(\bar L\log(\bar
L/\epsilon)/\epsilon)/\Delta) = \cO(\log(\bar L/\epsilon)/Delta)$ uses
of unitaries for an overall error bounded by $\epsilon$. 
}

The most salient complexities are summarized in
Table~\ref{table:summary}. It is worth noting that many applications
of AST to search and optimization problems satisfy that $L=O(1)$, in
which case the complexity is dominated by $1/\Delta$ with additional
logarithmic factors in $1/\epsilon$ and $v_{\rm max}/v_{\rm avg}$. For
example, this holds in the direct application to Grover's search
problem, where $H(s)$ linearly interpolates between projectors onto the
initial and final states, respectively. In this case we have
sufficient knowledge of the eigenphases.
If we apply our methods without using knowledge of the local behavior
of the gap or the rate of change of the states, then our complexities
have the expected quadratic speedup over classical search except for a
logarithmic factor due to the speedup of the path near where the gap
is minimal. However, in this case we are lucky: The maximum speed is
attained in the exact middle of the path, which is at an angular
distance of almost $\pi/4$ from either end. Because the algorithms use equal
subdivision to recursively implement the state transformation
(Sect.~\ref{sec:transformation_along_a_path}), the transformation
succeeds more quickly than expected, and the extra logarithmic factor
is dropped. For applications to search and optimization there is no
reason to transform more than one copy.

After we outline the conventions used in this paper, we introduce a
number of basic oracles that encapsulate the elementary operations
that we need to perform the state transformations. The oracles are
defined to be error-free. Their actual implementations in terms of the
$U_s$ are based on standard phase-estimation techniques with
well understood error behavior. In
Sect.~\ref{sec:state_transformation} we develop several procedures for
transforming an input state $\ket{\psi}$ into $\ket{\phi}$ in one step.
The procedures depend on how much is known about the two states and
their overlap. In Sect.~\ref{sec:transformation_along_a_path} we
compose these steps for transformations along a path. When overlaps
are not known, this requires an analysis of a recursively defined tree
of intervals, which we perform in sufficient detail to enable cost
estimates that are sensitive to variations in the gap $\Delta_s$ along
the path. Most of our algorithms are described with components whose
number of steps is random and the primary complexity given is the
average number of steps. To ensure that reversible versions of our
algorithms can be constructed with no change in complexity,
we keep track of the tail behavior of the number of steps, which
always has an exponential decay. The reversible implementations have a
built-in deterministic stopping criterion that is a multiple of the
average. Sect.~\ref{sec:sd} summarizes the complexities of our state
transformation algorithms in a table and considers how our results can
be generalized to the situation where the eigenphases are degenerate,
and the goal is to transform a state in an eigenspace of $U_0$ into some
unspecified state in a corresponding eigenspace of $U_1$.


\section{Conventions}
\label{sec:conventions}

Kets are normalized states unless explicitly stated otherwise. When
writing states such as $\ket{\varphi}$ for real numbers $\varphi$, we
assume that $\ket{\varphi}$ are orthonormal states for distinct
$\varphi$. The numbers $\varphi$ are to be expressed in terms of
labels of computational basis states for a finite system in a
reasonable way. For instance, $\varphi$ could be written in binary,
with the digits corresponding to basis states of qubits. The precision
used should be appropriate for the context. 

When writing linear expressions involving eigenphases, we always
intend them to be valid modulo $2\pi$. For example, when constraining
an eigenphase $\varphi$ by $\varphi \in
[\varphi_0-\delta,\varphi_0+\delta]$, it is intended to be read with
the expression ``${\rm mod}(2\pi)$'' appended.

For a number of parameters (such as $\Delta$, $L$, eigenphases and
error bounds), we require that they have ``reasonable'' values. For
example, eigenphase gaps should be less than $\pi$ and error bounds
less than $1$. We normally take such constraints for granted without
specifying them explicitly. In most cases, it suffices to replace the
parameter by a nearby sensible value if the parameter is out of
range.

To transform states along an eigenpath of a path of unitary operators,
we ultimately use controlled forms of these unitary operators and
their inverses. However, we initially provide algorithms calling on
idealized operators defined in terms of the unitary operators and
their eigenstates of interest. We refer to these operators as oracles.
When analyzing the complexity of algorithms, we do not distinguish
between controlled and uncontrolled applications of unitary operators
or oracles and their inverses. In general, we do not explicitly
mention the adjective ``controlled'' or the term ``inverse'', leaving
them implied.

When specifying the behavior of an oracle or subroutine, we use the
expression ``combination of states'' to refer to any superposition
and/or mixture of the given states. Formally, a combination of the
states $\ket{t_i}$ is a state of the form
$\sum_i\ket{t_i}\ket{e_i}_E$, where $\ket{e_i}$ are unnormalized
states of $E$ and $E$ is a system independent of previously introduced
ones. When we say that $T$ is an operator that transforms $\ket{\psi}$
into a combination of the states $\ket{t_{\psi,i}}$, we mean that
$T(\ket{\psi}) = \sum_i\ket{t_{\psi,i}}\ket{e_{\psi,i}}_E$, where $E$
is a system introduced by $T$ and $\ket{e_{\psi,i}}$ are unnormalized
states of $E$. The total state associated with the combination is
normalized. We always define such operators so that orthogonal states
are transformed into distinguishable combinations, and require $T$ to act
isometrically. The implicitly introduced system $E$ is different for
each instance of $T$ in an algorithm. The particulars of the
combination may also vary with instance. Thus, if an operator or
oracle $T$ has been defined in terms of combinations of states for a
family of input states, the symbol $T$ refers to an arbitrary operator
satisfying the definition each time it is used. Formally, one can
achieve this effect by transforming expressions as follows: For each
occurrence of $T$ replace it with an occurrence-specific new symbol
$T'$ and prefix the expression with ``for some operator $T'$
satisfying the definition of $T$''.

We intend $T$ to be reversible. In the absence of true decoherence
processes, this can always be achieved. We may use semi-classical
language to describe various computational actions such as setting a
newly introduced register to a particular state, but implicitly rely
on such actions having reversible forms. If the reverse of $T$
immediately follows $T$, the system $E$ is effectively eliminated by
being returned to an initial state. If we performed some other action
before reversing $T$, $E$ may play a decohering role. Note that since
$E$ is implicit in the definition of $T$, it is not accessible to
actions not involving $T$. If reversals are used, which instance of
$T$ is reversed needs to be stated if it is not clear from context.
When reversals are not used, the statement that $T$ maps $\ket{\psi}$ to a
combination of states $\ket{t_{\psi,i}}$ is equivalent to the
statement that $T$ is a quantum operation satisfying that the support
of $T(\ketbra{\psi}{})$ is in the span of the $\ket{t_{\psi,i}}$.
Besides allowing for reversals of instances of $T$, defining
combinations in terms of implicit systems enables amplitude-based
error bounds.

We may want to compare $T$ to an implementation $W$. Suppose $T$ has
been defined on input states $\ket{\psi}$ as above. We say that the
error amplitude of $W$ with respect to $T$ is $a$ if $W$ transforms
the input states into the specified combinations up to a term of
absolute amplitude at most $a$. (We omit the adjective ``absolute'' if
it is clear from context.) That is, $W(\ket{\psi}) =
\sum_i\ket{t_{\psi,i}}\ket{f_{\psi,i}}_F + \ket{r_{\psi}}$, where
$\|\ket{r_{\psi}}\| \leq a$. Note that the error amplitude is defined as
an upper bound, not an exact error or distance. We use the fact that
error amplitudes are subadditive under composition of in-principle
reversible processes. Specifically, this holds whenever the processes
can be realized unitarily by addition of ancillary systems. This may
require replacing explicit measurements by steps that reversibly
record the measurement outcome in the ancillary systems, and
performing steps that are conditioned on previous measurement outcomes
by the appropriate quantum-controlled steps. After this change,
subadditivity of the error amplitudes follows by writing the final
state as a combination of the error free state and the errors
introduced by each step propagated through the subsequent ones.
Observe that subsequent steps do not change the amplitude of the
propagated error.

Most of the procedures that we analyze are not explicitly formulated
in reversible form, and the primary complexity measure is an average
cost. Because state transformation procedures have applications as
subroutines in larger quantum algorithms, it is desirable to have
reversible versions that can exploit quantum parallelism without
introducing unwanted decoherence. However, the average cost is
determined with respect to stopping criteria associated with
measurements. When the procedure is reversified, one cannot have a
stopping criterion that is input dependent. In principle, this may
require running the procedure much longer than suggested by the
average cost to ensure that all possible computation paths terminate.
Suppose the average cost is $\bar C$. If we allow for some error
amplitude, we can set an absolute termination criterion by stopping
when the total cost has exceeded $C_{{\rm max}}$. This can be done in
a reversible way and introduces an error amplitude bounded by
$\sqrt{\bar C/C_{{\rm max}}}$ (Markov's inequality for non-negative
random variables converted to amplitude). This bound is undesirably
large and, without additional assumptions, can be approached. So we
seek procedures where the probability distribution of $C$ decays
exponentially after some multiple of $\bar C$. If ${\rm Prob}(C\geq
c)\leq x^{c-\lambda \bar C}$ for some $0<x<1$ and $\lambda\geq 1$,
then the error amplitude for $C_{{\rm max}} >\lambda \bar C$ is
bounded by $x^{(C_{{\rm max}}-\lambda\bar C)/2}$. This implies that
for an error amplitude of $\epsilon$, we can set $C_{{\rm max}} =
\lambda\bar C+2\ln(1/\epsilon)/\ln(1/x)$, so that the error dependence
of the cost has an additive term that is only logarithmic in
$\epsilon$.

In order to keep track of the exponential decay of costs, we use the
large-deviation technique of bounding the expectation
$\langle\Gamma^{C}\rangle$ of $\Gamma^{C}$. Thus, whenever it matters,
we specify the tail behavior of the probability distribution of $C$ by
an inequality of the form $\langle\Gamma^{C}\rangle\leq \Gamma^{\tilde
C}$ for $1\leq\Gamma<\Gamma_{{\rm max}}$. Rather than trying to
optimize the inequality, we generally choose convenient, simple
expressions for $\Gamma_{{\rm max}}$ and $\tilde C$, ensuring that
$\tilde C$ is bounded by a constant multiple of the average cost. This
suffices for stating bounds on complexities while having reasonable
estimates for the hidden constants. Given such a bound, we can use
Markov's inequality to show that ${\rm Prob}(C\geq c) = {\rm
Prob}(\Gamma^C\geq \Gamma^c) \leq \Gamma^{\tilde C-c}$. This is of the
desired form, with $x=1/\Gamma$. When proving tail bounds, we liberally use the
fact that $F(\lambda) = \langle e^{\lambda C}\rangle$ is log-convex in
$\lambda$. In particular, if $\langle\Gamma_1^{C}\rangle \leq
\Gamma_1^{\tilde C}$ for some $\Gamma_1\geq 1$, then this inequality
automatically holds for all $\Gamma$ between $1$ and $\Gamma_1$.

The main reason to use the large-deviation technique of the previous
paragraph is to simplify the estimation of tail bounds for total
costs of compositions of procedures. For this purpose we
have the following lemmas:

\begin{lem}\label{lem:ldev1}
Consider a sequence of procedures $S_j$ with costs
$C_j$ satisfying $\langle \Gamma^{C_j}|C_1,\ldots,C_{j-1}\rangle \leq
\Gamma^{\tilde C_j}$ for $1\leq \Gamma\leq \Gamma_{{\rm max}}$. Define
$C_{{\rm tot},l} = \sum_{i=1}^l C_i$ and $\tilde C_{{\rm tot},l} =
\sum_{i=1}^l \tilde C_i$. Then $\langle \Gamma^{C_{{\rm tot},l}}\rangle
\leq \Gamma^{\tilde C_{{\rm tot},l}}$ for $1\leq \Gamma\leq
\Gamma_{{\rm max}}$.
\end{lem}

For random variables $A$ and $B$, the expression $\langle A|B\rangle$
used in the lemma denotes the conditional expectation of $A$ given
$B$.

\begin{proof}
The proof is by induction on $l$ using a standard large-deviations
approach. Let $\mu$ denote the measure for the probability
distribution of its arguments.
\begin{eqnarray*}
\langle\Gamma^{C_{{\rm tot},l+1}}\rangle 
  &=& \int
  \langle\Gamma^{C_{l+1}}|C_1,C_2,\ldots C_l\rangle
    \Gamma^{C_{{\rm tot},l}}d\mu(C_1,\ldots ,C_l) \\
  &\leq& \Gamma^{\tilde C_{l+1}}\int 
    \Gamma^{C_{{\rm tot},l}}d\mu(C_1,\ldots ,C_l) \\
  &\leq& \Gamma^{\tilde C_{l+1}} \Gamma^{\tilde C_{{\rm tot},l}} \\
  &=& \Gamma^{\tilde C_{{\rm tot},l+1}}\;.
\end{eqnarray*}
\end{proof}

Lemma~\ref{lem:ldev1} implies that if each component procedure has
exponentially decaying cost above a multiple of the average, so does
the composition. The next lemma generalizes this result to the case where the number of $S_j$
invoked is not deterministic. In the lemma's statement, the binary
random variable $W_j$ can be thought of as ``$S_j$ was successful''.
The lemma is intended to be applied when $S_k$ depends only on which
of the $S_j$ with $j<k$ where successful. For a sequence of random
variables $X_i$, we write $\mathbf{X}_i=(X_1,\ldots,X_i)$.

\begin{lem}\label{lem:ldev2}
Consider a sequence of procedures $S_j$ with costs $C_j$. Let $W_j$ be
a binary random variable such that $C_j$ and $W_j$ are conditionally
independent of $\mathbf{C}_{j-1}$ given $\mathbf{W}_{j-1}$. Let
$V_j=1$ if $C_j>0$ and $V_j=0$ otherwise. Define $m=\sum_j V_j$ and
suppose that $\langle \Lambda^m\rangle \leq \Lambda^{\tilde m}$ for
$1\leq \Lambda\leq \Lambda_{{\rm max}}$ and $\langle
\Gamma^{C_j}|\mathbf{W}_j,V_j\rangle \leq \Gamma^{\tilde C}$ for $1\leq
\Gamma\leq \Gamma_{{\rm max}}$. Then $\langle
\Gamma^{C_{{\rm tot},\infty}}\rangle \leq \Gamma^{\tilde m\tilde C}$
for $1\leq\Gamma\leq\min\left(\Lambda_{{\rm max}}^{1/\tilde
C},\Gamma_{{\rm max}}\right)$.
\end{lem}
The proof is given in Appendix~\ref{app:pldev2}.


\section{Oracles}
\label{sec:phase_estimation_oracles}

We introduce four oracles implementing idealized quantum operations.
We can implement versions of these oracles with low error amplitude in
terms of subroutines with full access to the unitary operators
defining the eigenpath.

We say that an oracle is $U$-controlled if it commutes with any $X$
that commutes with $U$ and acts on the same system as $U$.  Such an
oracle necessarily preserves eigenstates $\ket{\psi}$ of $U$ in the
sense that it transforms $\ket{\psi}\ket{a}$ to a state of the form
$\ket{\psi}\ket{f(a,\psi)}$ for any state $\ket{a}$ of other systems.
In particular, if an instance of the oracle is reversed on a state of
the form $\ket{\psi}\ket{b}$ where $\ket{\psi}$ is an eigenstate of
$U$ and $\ket{b}$ is arbitrary, the result is of the form
$\ket{\psi}\ket{a'}$.  A strictly $U$-controlled oracle is obtained if
it is implemented solely in terms of controlled-$U$ operations.  The
implementations described for oracles below that are required to be
$U$-controlled satisfy this property.

Our oracle implementations have a specified error amplitude but are
still strictly $U$-controlled when this is required.  It is worth
noting that for strictly $U$-controlled implementations, if a bound on
the error amplitude is specified only for eigenstates of $U$, then it
holds in general. To see that it holds for arbitrary superpositions of
eigenstates, it is necessary to use the fact that the error amplitudes
for orthogonal eigenstates are orthogonal.  This holds because of
being $U$-controlled: For eigenstate inputs, the decomposition of the
output state into an error-free part and the error amplitude results
in both being a product of the eigenstate with states of other
systems.  We note that in general, given error amplitudes only for a
basis of the state space, errors in a superposition can add as
amplitudes rather than probabilities. This may result in a $\sqrt{d}$
error enhancement, where $d$ is the dimension.

\begin{dfn}\label{dfn:phase_estimation_oracle}
Given a unitary operator $U$ and resolution
$\delta$, a phase estimation oracle ${\rm PE}(U,\delta)$ is an
isometry that transforms eigenstates $\ket{\psi}$ of $U$ with
eigenphase $\varphi$ into a combination of states of the form
\begin{equation}
  \ket{\psi}\kets{\varphi_x}{A}\;, \end{equation} where
$\varphi_x-\varphi \in [-\delta, \delta]$. We require that ${\rm PE}$
is $U$-controlled. We can implement ${\rm PE}$ with error amplitude
$\epsilon$ using $\cO(\log(1/\epsilon)/\delta)$ applications of $U$.
\end{dfn}

To implement ${\rm PE}$ it suffices to use a high-confidence version
of phase estimation. Such a phase estimation technique and its
analysis are in Ref.~\cite{knill_expectation_2007}. ${\rm PE}$ is
$U$-controlled because all actions involving $U$'s system are
ancilla-controlled $U$ operations.

\begin{dfn}\label{dfn:phase_detection_oracle}
Given a unitary operator $U$, a phase $\varphi_0$ and
a resolution $\delta$, a phase detection oracle
${\rm PD}(U,\varphi_0,\delta)$ is an isometry that acts on
eigenstates $\ket{\psi}$ of $U$ with
eigenphase $\varphi\in [\varphi_0-\delta/4,\varphi_0+\delta/4]$
or $\varphi\not\in[\varphi_0-3\delta/4,\varphi_0+3\delta/4]$
as
\begin{equation}
{\rm PD}(U,\varphi_0,\delta)\ket{\psi}
  = \ket{\psi}\kets{b}A\;,
\label{eq:pdo_definition}
\end{equation}
where $b=1$ if $\varphi\in[\varphi_0-\delta/4,\varphi_0+\delta/4]$ and
$b=0$ otherwise. Eigenstates $\ket{\psi}$ not satisfying the
eigenphase constraint are mapped to arbitrary combinations of states
of the form $\ket{\psi}\kets{b}{A}$. We require that ${\rm PD}$ is
$U$-controlled. We can implement ${\rm PD}$ with error amplitude
$\epsilon$ using $\cO(\log(1/\epsilon)/\delta)$ applications of $U$.
\end{dfn}

To implement the phase detection oracle with the stated complexity,
apply ${\rm PE}(U,\delta/4)$, labeling its output register $A'$.
Reversibly set register $A$ to $\ket{1}$ if
$\varphi_x\in[\varphi_0-\delta/2,\varphi_0+\delta/2]$, otherwise set it to
$\ket{0}$. Then reverse the instance of ${\rm PE}$ used.

\begin{dfn}\label{dfn:reflection_oracle}
Given a unitary operator $U$, a phase $\varphi_0$ and a resolution
$\delta$, a reflection oracle ${\rm R}(U,\varphi_0, \delta)$ is an
isometry that acts on $\ket{\psi}$ in the subspace spanned by
eigenstates of $U$ with eigenphase $\varphi\in
[\varphi_0-\delta/4,\varphi_0+\delta/4]$ or
$\varphi\not\in[\varphi_0-3\delta/4,\varphi_0+3\delta/4]$ as
\begin{equation}
{\rm R}(U,\varphi_0,\delta)\ket{\psi}
  = (-1)^b\ket{\psi}\;,
\end{equation}
where $b=1$ if $\varphi\in[\varphi_0-\delta/4,\varphi_0+\delta/4]$ and
$b=0$ otherwise. Eigenstates $\ket{\psi}$ not satisfying the
eigenphase constraint are mapped to a combination of $\ket{\psi}$. We
require that ${\rm R}$ is $U$-controlled. We can implement ${\rm R}$
with error amplitude $\epsilon$ using $\cO(\log(1/\epsilon)/\delta)$
applications of $U$.
\end{dfn}

To implement the reflection oracle with the stated complexity, apply
${\rm PD}(U,\varphi_0,\delta)$. Conditionally on the bit in register
$A$ change the phase of the input state. Then reverse the instance of
${\rm PD}$ used.

A reflection oracle preserves eigenstates but can decohere the phases
for eigenstates not satisfying the constraint by correlating them with
the instance-dependent system implied by our convention for
combinations.

We note that the implementation of the reflection oracle is a special
case of the functional calculus of $U$ implemented via phase
estimation. For functions $f$ preserving the unit circle in the
complex plane and slowly varying except in known eigenphase gaps, one
can implement $f(U)$ by using phase estimation with sufficient
resolution, changing the phase according to $f$ and reversing the
phase estimation. The error can be bounded in terms of the
resolution used. A version of this observation for more general $f$
(which requires postselection) but not using high-confidence phase
estimation can be found in Ref.~\cite{harrow:qc2008a}.

\begin{dfn}\label{dfn:overlap_oracle}
Given states $\ket{\psi}$, $\ket{\phi}$, an overlap
threshold $\alpha$ and a resolution $\delta$,
an overlap detection oracle ${\rm OV}(\psi,\phi,\alpha,\delta)$ is an
isometry that transforms $\ket{\psi}$ into a combination of states of
the form
\begin{equation}
  \ket{\psi'}\kets{b}{A}\;,
\end{equation}
where the following holds: If
$\arccos(|\braket{\psi}{\phi}|)<\alpha-\delta$, then $b=1$ and
$\ket{\psi'}=\ket{\psi}$.  If
$\arccos(|\braket{\psi}{\phi}|)>\alpha+\delta$, then $b=0$ and
$\ket{\psi'}=\ket{\psi}$. Otherwise, $\ket{\psi'}$ is in the subspace
spanned by $\ket{\psi}$ and $\ket{\phi}$.  We can implement ${\rm
  OV}(\psi,\phi,\alpha,\delta)$ with error amplitude $\epsilon$ using
$\cO(\log(1/\epsilon)/\delta)$ applications of reflections around
$\ket{\psi}$ and $\ket{\phi}$.
\end{dfn}

Let $R_\psi$ and $R_\phi$ be reflections around $\ket{\psi}$ and
$\ket{\phi}$, respectively. The overlap oracle is implemented by
applying the phase estimation oracle ${\rm PE}(U=R_\psi
R_\phi,2\delta)$. The operator $U$ preserves the subspace spanned by
$\ket{\psi}$ and $\ket{\phi}$, and its two eigenvalues on this
subspace are $e^{\pm i 2 \arccos(|\braket{\psi}{\phi}|)}$. We set
register $A$ to $\ket{1}$ if the phase $\varphi_x$ returned by the phase
estimation oracle satisfies $\varphi_x \in[-2\alpha,2\alpha]$, and to
$\ket{0}$ otherwise. We then reverse the instance of ${\rm PE}$ used.
See Ref.~\cite{knill_expectation_2007} for more details.

If the reflections required for the overlap oracle are implemented in
terms of reflection oracles with resolution $\delta'$, the overall complexity
for an error amplitude of $\epsilon$ has a factor of
$\log(1/\epsilon)^2$. We generally aim for a dependence on $\epsilon$
of $\log(1/\epsilon)$, which requires bypassing direct uses of overlap
oracles.

\section{One-step state transformations}
\label{sec:state_transformation}

\begin{figure}[ht!]
\begin{align*}
  \Qcircuit @C=1em @R=1em { 
\lstick{\ket{0}} &\gate{H} & \ctrl{1} &  \qw & \gate{H} &\meter & \cw \\ 
 &\gate{R_{\psi}}& \gate{R_\phi} &  \qw &\qw
} 
\end{align*}
\caption{ Quantum circuit ${\rm RT}(\psi,\phi)$ for a state
  transformation attempt. $H$ denotes the Hadamard gate. The filled
  circle denotes control on the state $\ket{1}$ of the corresponding
  ancilla qubit.}
\label{fig:transform_by_reflection}
\end{figure}
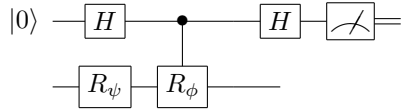

Suppose we are given a system in the eigenstate $\ket{\psi}$ of $U$ and
we wish to transform $\ket{\psi}$ into the eigenstate $\ket{\phi}$ of $V$.
We denote the eigenphases by $\varphi_U$ and $\varphi_V$,
respectively. We assume that the eigenstates are unique for the
eigenphases and that the gaps to the nearest other eigenphases are
bounded below by $\Delta$. The overlap probability is denoted by
$p=|\braket{\psi}{\phi}|^2$. Define $q=(1-p)$. The methods for
transforming the states depend on what is known about the eigenphases,
gaps and overlaps. Given reflection oracles and $p$ bounded away
from $0$ and $1$, we can use ideas developed for fixed point quantum
search~\cite{tulsi_fixedpoint_2006} and QMA
amplification~\cite{marriott_quantum_2005,nagaj_fast_2009} for a
transformation using a constant number of reflections on average.
Define reflection operators by $R_{\psi}=\one -2 \ket{\psi}\bra{\psi}$
and $R_{\phi}=\one -2 \ket{\phi}\bra{\phi}$. Each reflection operator
can be implemented with one call to the appropriate reflection
oracle. The transformation from $\ket{\psi}$ into $\ket{\phi}$ is
accomplished by repeatedly applying the circuit
${\rm RT}(\psi,\phi)$ of Fig.~\ref{fig:transform_by_reflection}
until the measurement outcome is $1$, indicating that $\ket{\phi}$ has
been prepared.

\begin{figure}
  \centering
  \includegraphics[width=7cm]{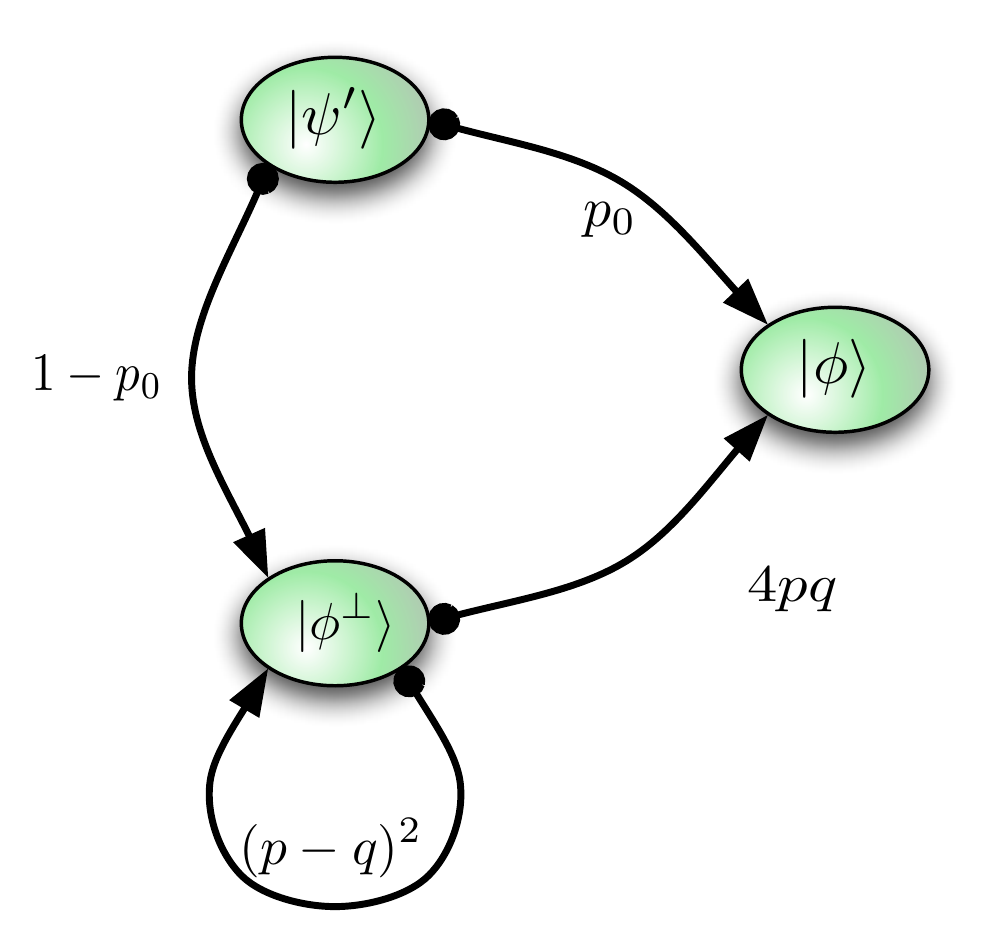}
  \caption{State diagram for $T(\psi,\phi)$. The state $\ket{\psi'}$
    is the initial state and $p_0$ is the overlap probability of
    $\ket{\psi'}$ with $\ket{\phi}$. For the non-trivial case of
    $p<1$, the transition probabilities from $\ket{\phi^\perp}$ are
    obtained by explicit computation of the reflections in the two
    dimensional subspace of the states. For this purpose, one can
    write $\ket{\psi}=\sqrt{p}\ket{\phi}+\sqrt{q}\ket{\phi^\perp}$,
    $\ket{\psi^\perp}=\sqrt{q}\ket{\phi}-\sqrt{p}\ket{\phi^\perp}$ and
    $\ket{\phi^\perp}=\sqrt{q}\ket{\psi}-\sqrt{p}\ket{\psi^\perp}$.}
    \label{fig:markov_chain_for_transform_by_reflection}
\ignore{
Start with $\ket{\phi^\perp}$ expressed in terms of $\ket{\psi}$
and $\ket{\psi^\perp}$: After reflection around $\ket{\psi}$,
one gets the state $-\sqrt{q}\ket{\psi}-\sqrt{p}\ket{\psi^\perp}$.
The overlap probability with $\ket{\phi}$ is obtained by expanding
in terms of $\ket{\phi}$ and $\ket{\phi^\perp}$. The
amplitude of $\ket{\phi}$ is $-2\sqrt{qp}$.
}
\end{figure}

The effect of ${\rm RT}(\psi,\phi)$ is to apply a reflection around
$\ket{\psi}$ followed by a projection onto $\ket{\phi}$ if the
measurement outcome is $1$, or a projection onto the orthogonal
complement otherwise. The subspace spanned by $\ket{\psi}$ and
$\ket{\phi}$ is preserved by the process. Let $\ket{\phi^\perp}$ be
the state orthogonal to $\ket{\phi}$ in this subspace. If the subspace
is one-dimensional, choose any orthogonal state. The procedure for
transforming the states can be analyzed as a Markov chain on the
states $\ket{\psi}$, $\ket{\phi}$ and $\ket{\phi^\perp}$. We consider
a slightly more general procedure $T(\psi,\phi)$ that can be applied
to any initial state $\ket{\psi'}$ in the subspace spanned by
$\ket{\psi}$ and $\ket{\phi}$. Define $p_0=|\braket{\psi'}{\phi}|^2$.
The first step of the procedure consists of the circuit for
${\rm RT}(\psi,\phi)$ with the reflection around $\ket{\psi}$ omitted.
Next, ${\rm RT}(\psi,\phi)$ is applied until the
measurement outcome on the ancilla qubit is $1$, indicating that
$\ket{\phi}$ has been prepared. The transition probabilities for this
procedure are shown in
Fig.~\ref{fig:markov_chain_for_transform_by_reflection}.

In general, when we describe a step of a procedure as a measurement of
a state $\ket{\phi'}$, this is intended to be implemented by means of
a controlled reflection around $\ket{\phi'}$ as in the second part of
${\rm RT}$. The effect is a projection onto $\ket{\phi'}$ or the
orthogonal complement, depending on the measurement outcome.

To simplify the notation, we omit arguments of procedures such as $T$
and ${\rm RT}$ when they are sufficiently clear from context.  The
arguments are typically passed on to appropriate oracle calls. We may
therefore use any set of alternative arguments that are sufficient for
specifying these oracles.

\begin{lem}\label{lem:transform_by_reflection}
We can transform any state $\ket{\psi'}$ in the subspace spanned by
$\ket{\psi}$ and $\ket{\phi}$ into $\ket{\phi}$ with a procedure $T$
using $\langle n\rangle = p_0+(1-p_0)(1+1/(2pq))\leq 1+1/(2pq)$
reflections around the states on average. For $1\leq \Gamma <
1/|p-q|$, define $c$ by the equation $\Gamma^2 = (1-cpq)/(p-q)^2$.
We then have $\langle \Gamma^n\rangle = \Gamma (p_0+(1-p_0)
4\Gamma^2/c)\leq 4\Gamma^3/c$.
\end{lem}

\begin{proof}
The procedure is as described above. Consider the process of
Fig.~\ref{fig:markov_chain_for_transform_by_reflection}. Define $n'$
to be the number of reflections used after the first step, if the
first step resulted in state $\ket{\phi^\perp}$. We have $\langle n
\rangle = p_0+(1-p_0)(1+\langle n'\rangle)$. If the first step failed,
the second step can either succeed or return to $\ket{\phi^\perp}$. In
the first case, we used $n'=2$ reflections. In the second case, the
expected number of reflections yet to be used is again $\langle
n'\rangle$. This implies the equation $\langle n'\rangle = 4pq\cdot 2
+ (p-q)^2(2+\langle n'\rangle)$. Thus $\langle n'\rangle = 2/(4pq)$
and $\langle n\rangle = p_0+(1-p_0)(1+1/(2pq))$.

Similarly, we can calculate $\langle \Gamma^n\rangle$ by solving the
equations $\langle \Gamma^n\rangle = p_0\Gamma + (1-p_0)\Gamma \langle
\Gamma^{n'}\rangle$ and $\langle\Gamma^{n'}\rangle = 4pq\Gamma^2 +
(p-q)^2(\Gamma^2\langle\Gamma^{n'}\rangle)$. This gives
$\langle\Gamma^{n'}\rangle = 4pq\Gamma^2/(1-(p-q)^2\Gamma^2) =
4\Gamma^2/c$. The last inequality follows because $c\leq 4$.
\end{proof}



According to Lemma~\ref{lem:transform_by_reflection}, 
the number of reflections used satisfies an exponential decay with a
fixed base less than $1$ provided that $p$ is bounded away from both
$0$ and $1$. In order to obtain a better behaved transformation that
only requires $p$ to be bounded away from $0$, we use an
overlap-suppression trick to modify $T$. We first add an ancilla $A$
in the state $\sqrt{3/4}\ket{0}_A+\sqrt{1/4}\kets{1}{A}$. Write
$\ket{\tilde\psi} =
\ket{\psi}(\sqrt{3/4}\ket{0}_A+\sqrt{1/4}\kets{1}{A})$ and
$\ket{\tilde\phi}=\ket{\phi}\kets{0}{A}$. Reflections around
$\ket{\tilde\psi}$ and $\ket{\tilde\phi}$ in the extended statespace
can be implemented with properly controlled reflections around
$\ket{\psi}$ and $\ket{\phi}$. We can therefore transform
$\ket{\tilde\psi}$ into $\ket{\tilde\phi}$ using
$T(\tilde\psi,\tilde\phi)$ After we are done, we discard the ancilla
to extract the desired state $\ket{\phi}$. The overlap-suppression
trick leads to the following lemma:

\begin{lem}\label{lem:tbr_with_bounded_overlap}
If $p\geq 1/3$, we can transform $\ket{\psi}$ into $\ket{\phi}$ with a
procedure $T_m$ whose average number of reflections satisfies
$\langle n\rangle<4$. For $1\leq\Gamma\leq 7/4$, we have
$\langle\Gamma^n\rangle\leq \Gamma^{6}$.
\end{lem}

\begin{proof}
With the technique just described, the overlap probability is changed
from $p$ to $3p/4$. Given the assumed lower bound, this ranges from
$1/4$ to $3/4$. Here, $p_0=3p/4$. From
Lemma~\ref{lem:transform_by_reflection}, the average number of
reflections used is at most $1+1/(2(1/4)(3/4)) < 4$. 

To prove the bound on $\langle \Gamma^n \rangle$, it suffices to show
that for $\Gamma\leq 7/4$, $\Gamma^3 \ge 4/c$ and apply the last
inequality of Lemma~\ref{lem:transform_by_reflection}.  The function
$(p-q)^2/(pq)$ is symmetric in $p$ and $q$ and achieves its maximum of
$1/2$ for the allowed values of $p$ at $p=1/3$.  Using the inequality
$(1+x)^y\geq 1+yx$ for $y\geq 0$ and $x>-1$, we get
\begin{eqnarray*}
\Gamma^3 \ge  \Gamma^{\frac{2(p-q)^2}{pq}}
         = (1+(\Gamma^2-1))^{\frac{(p-q)^2}{pq}}
         \ge 1+\frac{(p-q)^2}{pq}(\Gamma^2-1).
\end{eqnarray*}
Since $ \Gamma^2 = (1-cpq)/(p-q)^2 =1+(4-c)pq/(p-q)^2$, we conclude
that $\Gamma^3 \ge 5-c$. The constraints on $\Gamma$ and $p$ imply
that $c \ge 1$, so $\Gamma^3 \ge 4\geq 4/c$.
\ignore{
\begin{verbatim}
% Octave:
% Formulas used here are symmetric in p and q.
% Calculate average number of reflections of Tm.
function avg = tm_avg(p)
  q = (1-p);
  avg = 1+(1/(2*p*q)); 
endfunction
tm_avg(1/8)
% = 5.5714
tm_avg(1/4)
% = 3.6667
%
% Calculate the bound on Gamma and the exponent.
function gexp = tm_gexp(p)
  q = (1-p);
  gexp = [sqrt((1-p*q)/((p-q)^2)), 3+2*((p-q)^2)/(p*q)];
endfunction
tm_gexp(1/8)
% 1.2583, 13.2857
tm_gexp(1/4)
% 1.8028, 5.6667
\end{verbatim}}
\end{proof}

The constants used for Lemma~\ref{lem:tbr_with_bounded_overlap} have
been chosen for convenience and have not been optimized. Changing the
lower bound on $p$ or the parameters of the ancilla state when
redefining $\ket{\psi}$ changes the bounds in the lemma but does not
affect the complexities to be derived later.

To deal with the problem of low overlap probability $p$, we extend
$T_m$ to a procedure $T_x$ with the property that the initial state
$\ket{\psi}$ is transformed if $p$ is large enough and unchanged
for $p$ too small.

\begin{lem}\label{lem:tbr_unconstrained}
We can implement a procedure $T_x$ that transforms $\ket{\psi}$ into a
combination of states of the form $\ket{\psi}\kets{0}{A}$ and
$\ket{\phi}\kets{1}{A}$, where register $A$ is $\ket{1}$ if $p>1/2$
and $\ket{0}$ if $p<1/3$. $T_x$ requires one overlap oracle call with
a resolution of $(\arccos(\sqrt{1/3})-\arccos(\sqrt{1/2}))/2$, a
reflection around $\ket{\psi}$, and, in the case where the state is
transformed into $\ket{\phi}$, one instance of $T_m$ with $p$ guaranteed
to be at least $1/3$.
\end{lem}

\begin{proof}
To implement $T_x$, we use the overlap oracle ${\rm
OV}(\psi,\phi,(\arccos(\sqrt{1/3})+\arccos(\sqrt{1/2}))/2,
(\arccos(\sqrt{1/3})-\arccos(\sqrt{1/2}))/2)$ once to obtain a
combination of $\ket{\psi'}\kets{b}{A}$ with $\ket{\psi'}$ in the span
of $\ket{\psi}$ and $\ket{\phi}$. We then use a reflection around
$\ket{\psi}$ to measure $\ket{\psi}$. If the state is determined to be
$\ket{\psi^\perp}$ or if $b=1$, we apply $T_m$. The properties of the
overlap oracle and the choice of parameters ensure that this
happens only if $p\geq 1/3$. It always happens if $p>1/2$.
\end{proof}

The procedure $T_x$ provides bounds on the overlap at the cost of
calling an overlap oracle. As noted in
Sect.~\ref{sec:phase_estimation_oracles}, this results in unwanted
overhead when accounting for error amplitudes. This can be avoided if
one does not seek guaranteed overlap bounds and accepts the
possibility that even at high overlap, the transformation may not
succeed.

\begin{lem}\label{lem:txprime}
We can implement a procedure $T_x'$ that transforms $\ket{\psi}$ into a
combination of states of the form $\ket{\psi}\kets{0}{A}$ and
$\ket{\phi}\kets{1}{A}$, where register $A$ is $\ket{1}$ with
probability $1-\frac{1-p}{1+p}(q-p)^4$, and the number of
reflections around $\ket{\psi}$ and $\ket{\phi}$ is bounded by $n\leq 5+n'$,
where $\langle n'\rangle\leq 1/(1-p)$ and $\langle\Gamma^{n'}\rangle
\leq (1-p)\Gamma/(1-p\Gamma)$ for $1\leq\Gamma<1/p$.
\end{lem}

Register $A$'s contents indicate whether the transformation succeeded.

\begin{proof}
To implement $T_x'$, we apply a $\ket\phi$-measurement followed by
at most two applications of ${\rm RT}$, stopping if $\ket{\phi}$
is detected in the measurement. If after this, the state is
$\ket{\phi^\perp}$ (as indicated by the last measurement), we
alternately make $\ket{\psi}$- and $\ket{\phi}$-measurements until
either $\ket{\psi}$ or $\ket{\phi}$ is detected. Register $A$ is set
to $\ket{1}$ if $\ket{\phi}$ was detected in the last measurement
made. The probability that register $A$ is $\ket{1}$ is at least the
probability that the first three steps of the process of
Fig.~\ref{fig:markov_chain_for_transform_by_reflection} terminate at
$\ket{\phi}$, which is $1-(1-p)(q-p)^4$ (see the proof of
Lemma~\ref{lem:transform_by_reflection}). We can improve this by
noting that conditional on the failure of these steps, the probability
of success in the sequence of alternating measurements is
$(1-p)p\sum_{k=}^{\infty} p^{2k}=p/(1+p)$. Multiplying by $(1-p)(q-p)^4$ and
adding to $1-(1-p)(q-p)^4$ we get the overall probability of
successful preparation of $\ket{\phi}$.

The first part of the procedure uses at most five reflections. The
probability of failure to produce an acceptable outcome in a given
measurement in the second part is $p$. Thus, if the first part fails,
the expected number of reflections used in the second part is
$1/(1-p)$. Setting $n'$ to the number of reflections used if the first
part fails (event $F$), we have $\langle\Gamma^{n'}|F\rangle = (1-p)\Gamma +
p\Gamma\langle\Gamma^{n'}|F\rangle$. To obtain the last statement of the
lemma, it suffices to solve this equation.
\end{proof}

For later use, we note the following refinement of the lemma:

\begin{cor}\label{cor:txprime_refinement}
When $T_x'$ is invoked on state $\ket{\psi}$, the distribution of the
number of reflections used by $T_x'$ and whether $T_x'$ succeeds is
independent of any previous events. We also have $\langle
\Gamma^{n'}|A\rangle\leq (1-p^2)\Gamma^2/(1-\Gamma^2 p^2)$ for
$1\leq\Gamma<1/p$, where $A$ indicates success or failure of $T_x'$.
\end{cor}

\begin{proof}
Let $E_\phi$ and $E_\psi$ be the events that the first part of
$T_x'$ fails and $\ket{\phi}$ or $\ket{\psi}$ are eventually obtained,
respectively. The probability of $E_\phi$ is $p/(1+p)$. Conditional on $E_\phi$, $n'$ is even, $n' \ge 2$, and we get
\begin{eqnarray*}
  \langle \Gamma^{n'}|E_\phi\rangle = \frac{1+p}{p} \sum_{n' \in\, \{2,4,\ldots\}} (1-p) p^{n'-1} \Gamma^{n'} = \frac{\Gamma^2
(1-p^2)}{1-\Gamma^2 p^2}
\end{eqnarray*}
Similarly, conditional on $E_\psi$, $n'$ is
odd, and $\langle\Gamma^{n'}|E_\psi\rangle = \Gamma
(1-p^2)/(1-\Gamma^2 p^2)$. The claim follows because $\Gamma \geq 1$.
\end{proof}

For concreteness, we give the following corollary:

\begin{cor}\label{cor:txprime}
For $p\geq 1/4$, $T_x'$ transforms into $\ket{\phi}\kets{1}{A}$ with
probability at least $19/20$. For $p\leq 3/4$, the average number of
reflections satisfies $\langle n\rangle \leq 9$, and 
$\langle\Gamma^{n}|A\rangle \leq \Gamma^{11}$ for
$1\leq\Gamma\leq 8/7$, where $A$ indicates success or failure of
$T_x'$.
\end{cor}

\begin{proof}
We use the lower bound $1-(1-p)(q-p)^4$ for the probability of
success. The minimum of $1-(1-p)(q-p)^4$ for $p\in[1/4,1]$ is at
$p=9/10$, and one can check that the value is above
$19/20$. Since $\Gamma\geq 1$, $f(p)=
(1-p^2)/(1-p^2\Gamma^2)$ achieves its maximum on $p\in[0,3/4]$ at
$p=3/4$. We show the last bound for $\Gamma=8/7$, which is sufficient by log-convexity. In this case $\Gamma^4 \ge f(3/4)$ and, since $n \le 5 + n'$, Cor.~\ref{cor:txprime_refinement} gives the bound.
\ignore{
The derivative of $(1-p)(1-2p)^n$ is
$(1-2p)^{n-1}(-(1-2p) - 2n(1-p))$, with critical point above $p=1/2$
at $p=(2n+1)/(2n+2)$. For $n=4$, this is $9/10$.
\begin{verbatim}
% Octave:
% Minimum probability of success above 1/2:
function psucc = tx_psucc(p)
  q = (1-p);
  psucc = [(1-(1-p)*(q-p)^4), (1-(1-p)*(q-p)^4/(1+p))];
endfunction
tx_psucc(9/10)
% = [0.95904, 0.97844]
tx_psucc(1/4)
% = [0.95312, 0.96250]
% Bound on Gamma and <Gamma^n>:
% Set p=3/4, Gamma = (4/3)(1-1/8) = 7/6.
function gfact = tx_gfact(p,g,k)
  gfact = [(1-p)/(1-p*g), (1-p^2)/(1-p^2*g^2), g^k];
endfunction
tx_gfact(3/4, 7/6, 5)
% 2.0, 1.8667, 2.1614
tx_gfact(3/4, 7/6, 4)
% 2.0, 1.8667, 1.8526
tx_gfact(3/4, 8/7, 4)
% 1.75, 1.6490, 1.7060
\end{verbatim}
}
\end{proof}

We let $T'_{mx}$ be the procedure obtained by combining the
overlap-suppression trick with $T'_x$. The procedures $T$, $T_m$,
$T_x$, $T'_x$ and $T'_{mx}$ can be implemented in terms of the
reflection oracles ${\rm R}$ when the states $\ket{\psi}$ and
$\ket{\phi}$ are specified to be eigenstates of $U$ and $V$ with
isolated eigenphases known to satisfy $\varphi_U\in
[\tilde\varphi_U-\Delta/4,\tilde\varphi_U+\Delta/4]$ and $\varphi_V\in
[\tilde\varphi_V-\Delta/4,\tilde\varphi_V+\Delta/4]$, and with gaps
lower bounded by $\Delta$. Here $\tilde\varphi_U$ and
$\tilde\varphi_V$ are assumed to be known, but not $\varphi_U$ and
$\varphi_V$. In this situation, we say that we know $\Delta/2$-ranges
for the eigenvalues. The instances of the reflection oracles used by
the procedures are given by ${\rm R}(U,\tilde\varphi_U,\Delta)$ and
${\rm R}(V,\tilde\varphi_V,\Delta)$. We choose the third parameter to
be as large as possible, because this decreases the complexity of
implementing the reflection oracles in terms of the unitary operators.
For the rest of this paper, whenever we use one of $T$, $T_m$ and
$T_x$, we assume that the reflections used are based on reflection
oracles ${\rm R}$ with the third argument given by a known lower bound
on the minimum gap, by default $\Delta$.

We are interested in the situation where the overlap probability $p$
is bounded away from $0$, but we do not know a sufficiently small
range for the eigenphase of $\ket{\phi}$ with respect to its defining
operator $V$ to use the requisite reflection oracle. We show that a
sufficiently small eigenphase range can be obtained with low error if
we transform many copies of $\ket{\psi}$ in parallel into $\ket{\phi}$,
provided $p$ is sufficiently large. The statement of the lemma
includes an optional projection $\Pi_0$ such that $\Pi_0 \ket \phi =
0$. This is later used in the context of the overlap-suppression trick
so the $0$ eigenvalue of the modified unitary does not get confused
with the original phases (for instance, see Lemma~\ref{lem:tbr_no_phase}). 

\begin{lem}\label{lem:parallel_eigenphase_range}
Suppose that $p>1/2+\gamma$ with $\gamma>0$, and $V=V_0\oplus \Pi_0$,
where $\Pi_0$ is a projector that we can use to control other
operations and $\Pi_0\ket{\phi} = 0$. Then, using $2r$
instances of phase estimation oracles ${\rm PE}(V,\Delta/5)$,
we can implement an isometry ${\rm ER}(V,\Pi_0,\Delta,\gamma)$ with the
following property up to an error amplitude of
$e^{-r\gamma^2}$: ${\rm ER}$ transforms
$\ket{\psi}^{\tensor r}$ into a combination of states of the form
\begin{equation}
 \ket{\phi}^{\tensor j}\ket{\phi^{\perp}}^{\tensor (r-j)}
   \ket{\varphi}_A\kets{j}{B}\;,
\end{equation}
where $j>r/2$ and $\varphi-\varphi_V\in[-\Delta/5, \Delta/5]$.
\end{lem}

\begin{proof}
Use system label $i$ for the $i$'th copy of $\ket{\psi}$. We can
modify the phase estimation oracles ${\rm PE}(V,\Delta/5)$ so that the
eigenphase register contains a non-eigenphase $\#$ if the input state
is in the support of $\Pi_0$. We implement ${\rm ER}$ by first using
these modified phase estimation oracles ${\rm PE}'$ independently on
each $\kets{\psi}{i}$, placing the eigenphase in register $A_i$. We
can express $\ket{\psi}$ as
$\ket{\psi}=\sqrt{p}\ket{\phi}+\sqrt{1-p}\ket{\phi^\perp}$. By
definition, the $i$'th instance of ${\rm PE}'$ acts as
$\ket{\phi}_i\mapsto \ket{\phi}_i\kets{w_i}{A_iE_i}$ for some
$\kets{w_i}{A_iE_i}$ and some instance-specific additional system
$E_i$. 

If we make conceptual $\ket{\phi}_i$-measurements, we detect
$\kets{\phi}{i}$ with probability $p>1/2+\gamma$. Thus, with
probability at least $1-e^{-2r\gamma^2}$ (Hoeffding's
inequality~\cite{hoeffding:qc1963a}), $j>r/2$ of the values
$\varphi_i$ in registers $A_i$ are in
$[\varphi_V-\Delta/5,\varphi_V+\Delta/5]$. Consider this case. Because
of the gap condition, other eigenphases are outside
$(\varphi_V-4\Delta/5,\varphi_V+4\Delta/5)$. Thus, there is a
$\varphi'$ with the property that more than half of the $\varphi_i$
are in $[\varphi'-\Delta/5,\varphi'+\Delta/5]$, and we are guaranteed
that $\varphi_V$ is within $2\Delta/5$ of $\varphi'$. Furthermore, any
$\varphi_i$ within $\Delta/5$ of $\varphi'$ is associated with
$\ket{\phi_i}$ and therefore within $\Delta/5$ of $\varphi_V$.

We do not make the measurements of the previous paragraph. Instead we
reversibly (and unitarily) determine whether an interval
$[\varphi'-\Delta/5,\varphi'+\Delta/5]$ containing more than half of
the $\varphi_i$ exists. Except for an error amplitude of
$e^{-r\gamma^2}$, the state satisfies the condition, which we now
assume. We reversibly compute the number $j$ and the median $\varphi$
of the $\varphi_i$ in the interval found. As discussed in the previous
paragraph, $j>r/2$ and $\varphi$ is within $\Delta/5$ of $\varphi_V$.
We place $j$ in register $B$ and $\varphi$ in register $A$ and reverse
all classical reversible computations that were required since
invoking the phase estimation oracles. Next we reorder systems so that
for $i\leq j$, the $i$'th group is in state
$\ket{\phi}_i\kets{w'_i}{A_iE_i}$ and for $i>j$, it is in the state
$\kets{\eta}{iA_iE_i} = {\rm PE}'\kets{\phi^\perp}{i}$. Here $\kets{w'_i}{A_iE_i}$ may be different
from $\kets{w_i}{A_iE_i}$ because of possible correlations with
$\varphi$. The states $\kets{\eta}{iA_iE_i}$ have not changed because
the eigenphases encoded in these states have no correlation with
$\varphi$.

For the last step of ${\rm ER}$, we reverse the appropriate instance of
${\rm PE}'$ on each register $\kets{\eta}{iA_iE_i}$. In order to do this we need to have
computed the reordering permutation into an additional register, which
we retain only if we need to reverse this instance of ${\rm ER}$. The
desired combination of states has now been obtained.
\end{proof}

From the proof of Lemma~\ref{lem:parallel_eigenphase_range}, it can be
seen that the error is such that the $r$ input systems' state remains
in the tensor product of the span of $\ket{\psi}$ and $\ket{\phi}$.

To complete the transformation from $\ket{\psi}^{\tensor r}$ to
$\ket{\phi}^{\tensor r}$ after applying ${\rm ER}$ without using an
excessive number of reflections when $p$ is close to $1$, we can use
the overlap-suppression trick.

\begin{lem}\label{lem:tbr_no_phase}
Suppose that $p> (4/3)(1/2+\gamma)$ with $\gamma>0$ given. Then there
is a procedure $T_p(U,V,\Delta,\gamma)$ that transforms
$\ket{\psi}^{\tensor r}$ into a combination of states of the form
$\ket{\phi}^{\tensor r}\kets{\varphi}{A}$ up to an error amplitude of
$e^{-r\gamma^2}$, where $\varphi-\varphi_V\in[-\Delta/5,\Delta/5]$.
$T_p$ uses less than $2r$ instances of ${\rm PE}(V,\Delta/5)$ and an average of
$\langle n\rangle<2r$ reflections. We have $\langle \Gamma^n\rangle
\leq \Gamma^{3r}$ for $1\leq\Gamma\leq 7/4$.
\end{lem}

\begin{proof}
We begin $T_p$ by adding ancilla qubits in state
$\sqrt{3/4}\ket{0}+\sqrt{1/4}\ket{1}$. Define $|\tilde\psi\rangle=
\ket{\psi}(\sqrt{3/4}\ket{0}+\sqrt{1/4}\ket{1})$ and
$\ket{\tilde\phi}=\ket{\phi}\ket{0}$. For the purposes of using ${\rm
  ER}$, let $\Pi_0=\one\tensor\ketbra{1}{}$, where the second factor
acts on the ancilla. We apply ${\rm
  ER}((V\tensor\ketbra{0}{})\oplus\Pi_0, \Pi_0,\Delta,\gamma)$ and,
provided $j>r/2$, we use $T(\tilde\psi,\tilde\phi)$ (defined
in~\ref{lem:transform_by_reflection}) on the registers whose state is
now indicated to be $\ket{{\tilde\phi}^\perp}$. The number of
instances of $T$ applied is less than $r/2$. The bounds on the number
of reflections used follow by
Lemmas~\ref{lem:tbr_with_bounded_overlap} and~\ref{lem:ldev1}.
\end{proof}

We need a method $T_{px}$ for transforming states as in
Lemma~\ref{lem:tbr_no_phase} that is well-behaved even for small $p$. In
order for the method to work we require that the probabilities of
eigenphases other than $\varphi_V$ in $\ket{\psi}$ have a boundedness
property. This will ensure that if $\ket{\psi}$ has a big overlap with an eigenstate of $V$, then this eigenstate is $\ket{\phi}$.

\begin{dfn}
\label{dfn:dominant}
We say that $\varphi_V$ is a $(\gamma,\delta)$-dominant eigenphase of
$V=V_0\oplus\Pi_0$ in $\ket{\psi}$ if for every $\varphi$ and
associated projector $\Pi$ onto eigenspaces of $V_0$ with eigenphases
in $I=[\varphi-\delta,\varphi+\delta]$, $|\Pi\ket{\psi}|^2 >
\gamma$ implies $\varphi_V\in I$. We refer to $\delta$ as the resolution
at which $\varphi_V$ is dominant and take $\Pi_0$ to be zero-dimensional
if it is not specified.
\end{dfn}

$T_{px}$ performs the transformation in the following steps.
The first determines whether it is possible to confidently find a
small interval for $\varphi_V$ without changing each of the $r$
copies of $\ket{\psi}$ by much. We then measure each copy so as to project
it onto $\ket{\psi}$ or $\ket{\psi^\perp}$. We ensure that the
probability of recovering a large number of copies of $\ket{\psi}$ is
high. If some $\ket{\psi^\perp}$ are found or if we found that
$\varphi_V$ can be determined sufficiently well, we use the recovered
copies of $\ket{\psi}$ to learn a small interval containing
$\varphi_V$ and then transform the states. The first step is
encapsulated by the next lemma.

\begin{lem}\label{lem:per_no_bound}
Suppose that $\varphi_V$ is a $(p_m-3\gamma,\delta)$-dominant
eigenphase of $V=V_0\oplus \Pi_0$ in $\ket{\psi}$, where $\Pi_0$ is a
projector that we can use to control other operations,
$\Pi_0\ket{\phi}=0$, $p_m-3\gamma> 1/2$ and $p_m< 1$. Let
$\delta'=\min(\delta/2,\Delta/4)$. Then, using $2r$ instances of phase
estimation oracles ${\rm PE}(V,\delta'/2)$ and $r$ reflections around
$\ket{\psi}$, we can implement an isometry
${\rm ER}_x(V,\Pi_0,\Delta,\delta,p_m,\gamma)$ with the following
property up to an error amplitude of $5e^{-r\gamma^2}$:
${\rm ER}_x$ transforms $\ket{\psi}^{\tensor r}$ into a combination of
states of the form
\begin{equation}
\ket{\psi}^{\tensor j}\ket{\psi^\perp}^{\tensor (r-j)}\kets{b}{A}\kets{j}{B}\;,
\label{eq:lem:per_no_bound}
\end{equation}
where if $p>p_m$, then $j=r$ and $b=1$;
if $p\leq p_m-2\gamma$, then $j=r$ and $b=0$; and 
otherwise $j\geq r/20$.
\end{lem}

\begin{proof}
We use the notation introduced in the proof of
Lemma~\ref{lem:parallel_eigenphase_range} and apply $r$ instances of
${\rm PE}'(V,\delta'/2)$ to accommodate the special subspace associated
with $\Pi_0$. We look (reversibly) for the first interval
$I_l=[(l-1)\delta',(l+1)\delta']$, $l\in
\{0,1,\ldots,\lceil 2\pi/\delta'\rceil-1\}$, containing at least
$(p_m-\gamma)r$ of the phases in registers $A_i$. If no such interval
exists, we set the state of $A$ to $\kets{0}{A}$, else we set it to
$\kets{1}{A}$. Any temporary storage required in the reversible
classical computation of the content of $A$ is erased. We then reverse
the instances of ${\rm PE}'$ used and make $\ket{\psi}$-measurements to
determine which of the $r$ input registers are in state $\ket{\psi}$.
Finally, we move the $j$ registers in this state to the front and set
the state of $B$ to $\kets{j}{B}$.

As in the proof of Lemma~\ref{lem:parallel_eigenphase_range}, after
phase estimation, for some set $S$ the state is a combination of
products of $\ket{\phi}_i\kets{w_i}{A_iE_i}$ for $i\in S$ and
$\kets{\eta}{iA_iE_i}$ for $i\not\in S$. There exists $l_0$ such that
$I_{l_0}\supset [\varphi_V-\delta'/2,\varphi_V+\delta'/2]$. Suppose
that we conceptually measure the registers $A_i$ before the reversal
of the phase estimation oracles. Let $k_l$ be the number of measured
phases that are in $I_l$. Because $2\delta'<\Delta$, $k_{l_0}=|S|$. In
particular, the measured phases in principle determine the members of
$S$. (We can use this for the analysis but not for the procedure.) We
consider the three cases $p>p_m$, $p\leq p_m-2\gamma$ and $p_m\geq
p>p_m-2\gamma$. First, if $p>p_m$, then, from Hoeffding's inequality
applied to $|S|$, the probability that $|S|>(p_m-\gamma)r$ is at least
$1-e^{-2r\gamma^2}$. Hence, with error amplitude at most
$e^{-r\gamma^2}$, there is a $k_l\geq(p_m-\gamma)r$, and register $A$
contains $1$ before the reversals of the phase estimation oracles. The
reversals successfully restore the initial state up to the given
error.

Consider next $p\leq p_m-2\gamma$. The probability that $|S|\geq
(p_m-\gamma)r$ is bounded by $e^{-2r\gamma^2}$.  We show that the
probability of finding a $k_l\geq (p_m-\gamma)r$ is small. For this
purpose, consider the set $H$ of $l$ such that $\varphi_V\not\in
I'_l=[(l-3/2)\delta',(l+3/2)\delta']$. For $l\in H$, any measured
phase in $I_l$ is associated with an eigenphase in $I'_l$ and
therefore different from $\varphi_V$. Because $(3/2)\delta'<\delta$,
the dominance condition ensures that such eigenphases occur with
probability at most $p_m-3\gamma$ in $\ket{\psi}$.  To obtain a good
bound on the mentioned probability, we consider the $k_l$'s according
to $l$'s location in a small partition of $H$.  For any $F\subseteq
H$, let $\varphi(F)$ be the set of eigenvalues of $V$ in $\bigcup
\{I'_l|l\in F\}$ and $P(F)$ the total probability of eigenphases in
$\varphi(F)$ in $\ket{\psi}$. We claim that we can in principle
partition $H=F_1\cup\ldots\cup F_{16}$ such that $P(F_i)\leq
p_m-3\gamma$. First note that $\sum_l P(\{l\})\leq 4$ because each
eigenphase occurs in at most $4$ of the $I'_l$. We can construct the
$F_i$ greedily.  Initialize $i=1$ and set $F_i=\emptyset$. Then step
through $l\in H$. If $P(F_i\cup\{l\})\leq p_m-3\gamma$, add $l$ to
$F_i$ and proceed to the next $l$. If not, because $p_m-3\gamma>1/2$,
either $P(F_i)>1/4$ or $P(\{l\})>1/4$. If $P(F_i)>1/4$, initialize
$F_{i+1}=\{l\}$, update $i$ to $i+1$ and proceed to the next $l$. If
$P(\{l\})>1/4$, set $F_{i+1}=F_{i}$, reset $F_{i}$ to $\{l\}$, update
$i$ to $i+1$ and proceed to the next $l$. At the end of the procedure,
$P(F_{i'})>1/4$ for all $i'<i$. As we observed above, for all $l\in H$,
$P(\{l\})\leq p_m-3\gamma$, and the claim follows. Let $k(F_j)$ be the
number of measured phases that are in $\bigcup_{l\in F_j}I_l$. Because
these phases are due to eigenphases in $\varphi(F_i)$, it follows from
Hoeffding's inequality that $k(F_i) \geq (p_m-\gamma)r$ with
probability at most $e^{-8r\gamma^2}$. If $l$ satisfies that
$\varphi_V\in I'_l$, then the measured phases in $I_l$ are associated
with $\varphi_V$. This is because $(7/2)\delta'<\Delta$. Probabilistic
reasoning can be applied, so we conclude that the probability of
$k_{{\rm max}}\geq (p_m-\gamma)r$ is at most $e^{-2r\gamma^2} + 16
e^{-8r\gamma^2} < 17 e^{-2r\gamma^2}$. Thus, with an error amplitude
of at most $5e^{-r\gamma^2}$, register $A$ contains $0$ before the
phase estimation reversals, and the reversals successfully restore the
initial state.

Finally, consider $p_m-\gamma\geq p>p_m-2\gamma$. Because $p_m -
3\gamma > 1/2$, the probability that $|S|>r/2$ is at least
$1-e^{-2r\gamma^2}$.  Therefore, except for an error amplitude of at
most $e^{-r\gamma^2}$, $k_{{\rm max}}>p-\gamma>p_m-3\gamma>1/2$ and
$k_{{\rm max}}=|S|$. We now assume this condition.  After reversing
the phase estimation oracles, the $i$'th system is in state
$\ket{\phi}$ if $i\in S$ and $\ket{\phi^\perp}$ otherwise. Let $j$ be
the number of $\ket{\psi}$ observed in the $\ket{\psi}$-measurements.
The probability of detecting $\ket{\psi}$ when the state is
$\ket{\phi}$ or $\ket{\phi^\perp}$ is $p$ and $1-p$, respectively. The
average value of $j$ is $p|S|+(1-p)(r-|S|)\geq r/2$, since $p>1/2$ and
$|S|>r/2$. By Hoeffding's inequality, for fixed $S$, the probability
of the event $E$ that $j\leq (1-x)r/2$ is bounded by $e^{-rx^2}$. To
determine a bound on the overall probability $P$ of $E$, we must use
amplitude addition over different $S$. Thus $\sqrt{P}\leq \sum_{S}
\sqrt{P_S}e^{-rx^2/2}$. There are $2^r$ possible $S$, so the worst
case sum of the $\sqrt{P_S}$ is $2^{r/2}$. Therefore, $P\leq
e^{r(\ln(2)-x^2)}$. We set $x=9/10$ and use $\ln(2)-(9/10)^2<-1/10$ to
see that the amplitude for having $j<r/20$ is bounded by
$e^{-r/20}$. By amplitude addition, the overall error amplitude is
bounded by $e^{-r\gamma^2}+e^{-r/20}<2 e^{-r\gamma^2}$, since
$\gamma<1/6$.

To complete the proof, it suffices to determine the maximum error
amplitude. The maximum error bound comes from the second case and is
given by $5e^{-r\gamma^2}$.
\end{proof}

As we noted for the error amplitude in
Lemma~\ref{lem:parallel_eigenphase_range}, the error in
Lemma~\ref{lem:per_no_bound} is such that the $r$ input systems' state
remains in the tensor product of the span of $\ket{\psi}$ and
$\ket{\phi}$.

The last lemma of this section gives the properties of the parallel
state transformation procedure $T_{px}$ that we outlined above.

\begin{lem}\label{lem:tbr_no_phase_unconstrained}
Assume that $\varphi_V$ is a $(1-4\gamma,\delta)$-dominant eigenphase
of $V$ in $\ket{\psi}$ with $1-4\gamma> 2/3$. Then there is a
procedure $T_{px}$ that transforms $\ket{\psi}^{\tensor r}$ into a
combination of $\ket{\phi}^{\tensor r}\kets{\varphi}{A}$ and
$\ket{\psi}^{\tensor r}\kets{\#}{A}$, where $\varphi-\varphi_V\in
[-\Delta/5,\Delta/5]$, $A$'s state is $\ket{\#}$ if $p\leq 1-3\gamma$
and $\ket{\varphi}$ if $p>1-\gamma$, and the error amplitude is
bounded by $6e^{-r\gamma^2/36}$. The procedure uses less than $4r$
instances of $\textrm{PE}(V,\delta'/2)$, where
$\delta'=\min(\delta/2,\Delta/4)$, and an average number of
reflections bounded by $\langle n\rangle\leq 5r$. We have
$\langle\Gamma^n\rangle\leq\Gamma^{7r}$ for $1\leq\Gamma\leq 7/4$.
\end{lem}

Note that the procedure implicitly provides overlap information. That is,
if the transformation succeeds, the overlap satisfies $p>1-3\gamma$.

\begin{proof}
We use the overlap suppression trick and change each copy of
$\ket{\psi}$ to $\ket{\tilde\psi} =
\ket{\psi}(\sqrt{3/4}\ket{0}+\sqrt{1/4}\ket{1})$ and define
$\ket{\tilde\phi} = \ket{\phi}\ket{0}$. Let
$\Pi_0=\one\tensor\ketbra{1}{}$. We apply the procedure
$\textrm{ER}_x(V\tensor\ketbra{0}{}\oplus\Pi_0,\Pi_0,\Delta,\delta,3(1-\gamma)/4,3\gamma/4)$
of the previous Lemma.
If the first output register (system A in
Eq.~\eqref{eq:lem:per_no_bound}) contains $b=1$ or if the second
register (system B in Eq.~\eqref{eq:lem:per_no_bound}) has $j<r$, we
continue. Otherwise we set the return register to $\ket{\#}$ and stop.

To continue the procedure, we apply $T_p(U,V,\Delta,(3/4)\gamma)$
(defined in~\ref{lem:tbr_no_phase}) to the first $j$ registers (that
now contain $\ket{\tilde\psi}$), omitting the initial
overlap-suppressing steps as they have already been done.  The
specification of $ER_x$ and the assumption $1-4\gamma > 2/3$ ensures
that the overlap is big enough to apply $T_p$, which returns
$\varphi$. We then apply the appropriate instances of $T$ (defined
in~\ref{lem:transform_by_reflection}) to the remaining $r-j$ registers
to transform $\ket{\tilde\psi^\perp}$ into $\ket{\tilde\phi}$. The
reflections around $\ket{\tilde\phi}$ implicitly require $\varphi$. To
finish we return the $r$ registers and $\kets{\varphi}{A}$.

The error amplitudes associated with the different steps must be
added. From the application of $\textrm{ER}_x$ we get
$5e^{-r(3/4)^2\gamma^2}$ to which $T_p$ adds at most
$e^{-(r/20)(3/4)^2\gamma^2}$. 

The number of instances of phase estimation oracles used comes from
the application of ${\rm ER}_x$ and $T_p$. The average number of
reflections is bounded by the sum of $r$ (from applying ${\rm ER}_x$),
$2j$ (from applying $T_p$), and at most $4(r-j)$ (from using
instances of $T$ as in Lemma~\ref{lem:tbr_with_bounded_overlap}).
The reflections in $T_p$ are from applications of at most $j/2$
instances of $T$. The total number of instances of $T$ is bounded
by $r$. To get the tail bounds,
apply the bounds from Lemmas~\ref{lem:tbr_with_bounded_overlap} and~\ref{lem:ldev1}, with an additional offset of $r$ for the
first set of reflections.
\end{proof}

\section{State transformations along a path}
\label{sec:transformation_along_a_path}

We consider paths of eigenstates $\ket{\psi_s}$ of unitary operators
$U_s$ with eigenphases $\varphi_s$ and gaps $\Delta_s$ as defined in
the introduction. We assume the ability to apply any $U_s$ and to
prepare $\ket{\psi_0}$. If the exact gaps are difficult to obtain, we
take the $\Delta_s$ to be known lower bounds on the gaps. We define
$p_{s,t}=|\braket{\psi_s}{\psi_t}|^2$. The goal is to transform copies
of the initial state $\ket{\psi_0}$ into the final state
$\ket{\psi_1}$.  The transformations' complexities depend on what is
known about the overlaps and the eigenphases along the path. They are
designed to provide such information if it is not already known, so
that future transformations can be performed more efficiently.

\begin{thm}\label{thm:simple_path}
Suppose that we know a subsequence $0=s_0<\ldots<s_n=1$ of $[0,1]$
such that $p_{s_k,s_{k+1}}\geq 1/3$, and phases $\tilde\varphi_{i}$
satisfying $\tilde\varphi_{i}-\varphi_{s_i}\in[-\Delta/4,\Delta/4]$.
We can then transform $\ket{\psi_0}$ to $\ket{\psi_1}$ with $m$
reflections ${\rm R}(U_{s_i},\tilde \varphi_i,\Delta_{s_i})$ where $\langle
m\rangle < 4n$ and $\langle\Gamma^m\rangle\leq \Gamma^{6n}$ for
$1\leq\Gamma\leq 7/4$.
\end{thm}

\begin{proof}
It suffices to apply $T_m$ with the reflections instantiated by
reflection oracles to advance from each state to the next. The
complexities follow from Lemma~\ref{lem:tbr_with_bounded_overlap} and~\ref{lem:ldev1}.
\end{proof}

Given sufficiently large overlaps, the phases can be inferred to
sufficient precision during a parallel state transformation. Note that
the eigenphase $\varphi_0$ of $\ket{\psi_0}$ for $U_0$ can be
determined to within $\Delta_0/4$ by one call to a phase estimation
oracle with resolution $\Delta/4$ and input state $\ket{\psi_0}$. We
therefore assume that a phase sufficiently close to $\varphi_0$ is
known and reflections around $\ket{\psi_0}$ can be applied.

\begin{thm}\label{thm:simple_path_no_phase}
Suppose that we know a subsequence $0=s_0<\ldots<s_n=1$ of $[0,1]$
such that $p_{s_k,s_{k+1}}>(4/3)(1/2+\gamma)$ with $\gamma>0$. We can
then transform $\ket{\psi_0}^{\tensor r}$ into $\ket{\psi_1}^{\tensor
r}$ with an error amplitude of $n e^{-r\gamma^2}$. The transformation
requires $2nr$ instances of phase estimation
${\rm PE}(U_{s_i},\Delta_{s_i}/5)$ and an average of $\langle m\rangle<
2nr$ reflections ${\rm R}(U_{s_i},\tilde \varphi_i,\Delta_{s_i})$. Furthermore
$\langle\Gamma^m\rangle\leq \Gamma^{3rn}$ for $1\leq\Gamma\leq 7/4$.
The transformation provides phases $\tilde\varphi_i$ satisfying
$\tilde\varphi_i-\varphi_{s_i}\in[-\Delta/5,\Delta/5]$ for $i>0$.
\end{thm}

\begin{proof}
It suffices to apply $T_p(U_{s_{i-1}},U_{s_i},\Delta_{s_i},\gamma)$ $n$ times to
transform the states. The complexities follow from
Lemma~\ref{lem:tbr_no_phase} and~\ref{lem:ldev1}.
\end{proof}

The number of underlying calls to phase estimation oracles per copy of
$\ket{\psi_1}$ produced by the procedure of
Thm.~\ref{thm:simple_path_no_phase} is within a constant factor of
that of Thm.~\ref{thm:simple_path} (where phase estimation is used for
the implementation of the reflections). The implementations of the
oracles in the former case have an additional overhead to achieve the
error goal, see Sect.~\ref{sec:sd}.

Theorems~\ref{thm:simple_path} and~\ref{thm:simple_path_no_phase}
suggest that we can obtain state transformations along paths with
complexities bounded by the path length. In particular, if the overlap
probabilities $p_{s_i,s_{i+1}}$ are bounded above by $\cos(\theta)^2$,
the path length is at least $n\theta$. Although it is possible for $n$
to be much smaller than the path length due to shortcuts, generically
we do not expect this. If the angular rate
$\chi(s)=\|(\one-\ketbras{\psi_s}{\psi_s}{})\ket{\partial_s\psi_s}\|$
along the path is defined and constant, $\chi(s)=\chi$, then regularly
spaced $s_i=\pi/(8\chi)$ ensure that the overlap conditions for the
theorems above are satisfied and $n=\lceil 8L/\pi\rceil$. On the
other hand, if many overlaps are close to $1$, $n$
could be large compared to $L$. We can eliminate this possibility if we have
sufficient information about the overlaps, or after the first
transformation by checking $n$ overlaps during the transformation, as
the next lemma shows.

\begin{lem}\label{lem:n_overlap_checks}
Let $\theta,\underline\theta,\overline\theta>0$ with
$\underline\theta<\overline\theta<\underline\theta+\theta<\pi/2$ be
given. Consider a procedure for transforming copies of $\ket{\psi_0}$
into copies of $\ket{\psi_1}$ in $n$ steps, where the steps transform
from $\ket{\psi_{t_{i-1}}}$ to $\ket{\psi_{t_i}}$ with
$\arccos(|\braket{\psi_{t_{i-1}}}{\psi_{t_i}}|)<\theta$. Neither $n$
nor the $t_i$ need to be deterministic, but we assume that after
$\ket{\psi_{t_i}}$ has been reached, information required to perform
reflections and call overlap oracles for states $\ket{\psi_{t_j}}$
with $j\leq i$ is available. We can then modify the procedure so that
it outputs a sequence $S=\{0=s_0<\ldots<s_k=1\}$ satisfying
$\arccos(|\braket{\psi_{s_{i-1}}}{\psi_{s_i}}|)\leq\overline\theta+\theta$
for all $i$ and
$\underline\theta\leq\arccos(|\braket{\psi_{s_{i-1}}}{\psi_{s_i}}|)$
for $i<k$ so that $L(S)\geq \underline\theta (k-1)$. To do so requires
$n$ calls to overlap oracles ${\rm
OV}(\ket{\psi_{t_l}},\ket{\psi_{t_j}},
(\overline\theta+\underline\theta)/2,
(\overline\theta-\underline\theta)/2)$, $n$ explicit reflections on $\ket{\psi_{t_l}}$ and $k$ invocations of
$T(\ket{\psi_{t_l}},\ket{\psi_{t_j}})$ with
$\cos(\overline\theta)^2\leq p\leq\cos(\underline\theta)^2$.
\end{lem}

The procedures we describe satisfy that the information required to
call reflection and overlap oracles is available when needed by the
modification in the lemma.

\begin{proof}
The $s_j$ are elements of $\{t_i\}_i$. We begin by setting $s_0=0$. We
ensure that at the end of the $l$'th step ($l\geq 1$) of the modified
procedure, the last $s_j$ that has been determined satisfies the
invariant
$\arccos(|\braket{\psi_{s_j}}{\psi_{t_{l}}}|)\leq\overline\theta$. To
do so, let $s_j$ be the last member of $S$ that has been determined
before the $l$'th step. If $l=n$, set $s_{j+1}=1$. Else, after the
transformation into copies of $\ket{\psi_{t_l}}$ has been accomplished,
call the overlap oracle ${\rm OV}(\ket{\psi_{t_l}},\ket{\psi_{s_j}},
(\overline\theta+\underline\theta)/2,
(\overline\theta-\underline\theta)/2)$ on the first copy of
$\ket{\psi_{t_l}}$. Then use a reflection around $\ket{\psi_{t_l}}$ to
determine whether the state was preserved. If not, or if the overlap
oracle returned $1$, set $s_{j+1} = t_l$. If the state was not
preserved, then call $T(\ket{\psi_{s_j}},\ket{\psi_{t_l}})$ to restore
it. This completes the modification of the $l$'th step and ensures the
desired properties for the $s_j$ determined so far and the invariant.
The lower bound on the length follows by adding up the lower bounds on
the angular distances between successive $\ket{\psi_{s_i}}$ for $i<k$.
\end{proof}

When the overlaps $p_{s,t}$ or the angular rates $\chi(s)$ are
unknown, the state transformation requires a recursive procedure to
find a sequence of successive states with sufficiently high overlap.
We assume that such states can be found, more specifically, we require
that there are no jumps of angular distance equal to some given
constant or greater. Our recursive state transformations involve binary
subdivision of intervals. To transform the state from $\ket{\psi_a}$
to $\ket{\psi_b}$, we check whether we can do it directly at a cost of
$C(a,b)$. If not, we recursively transform from $\ket{\psi_a}$ to
$\ket{\psi_{(a+b)/2}}$ and then from $\ket{\psi_{(a+b)/2}}$ to
$\ket{\psi_{b}}$. We are interested in the total cost of the
transformation. For our purposes, the cost is the number of times the
unitaries $U_s$ are used. This is determined by the number of times
phase-estimation is used either directly or indirectly when applying
reflections. The resolution required is typically the gap, and the
cost is related to the inverse gap
(Sect.~\ref{sec:phase_estimation_oracles}), which can depend on the
position along the path. To enable taking this into account we provide general tools
for analyzing the complexity of recursive path transformations based
on binary subdivision in Appendix~\ref{app:rtc}.

We define the symmetric binary interval tree on $[a,b]$, ${\rm
  BIT}(a,b)$, as the set of intervals constructed by starting with
$T=T_0=\{[a,b]\}$ and recursively adjoining $[c,(c+d)/2]$ and
$[(c+d)/2,d]$ to $T$ for every $[c,d]$ in $T$. We also define the cost
of ${\rm BIT}(a,b)$ as
\begin{equation}
C({\rm BIT}(a,b)) = 
  \sum_{[c,d]\in{\rm BIT}(a,b)} C(c,d)\;.
\end{equation}
With the appropriate choice of the cost function $C$, this is the cost
of a recursive state transformation procedure, and we show that it
depends linearly on length and at worst logarithmically on the ratio
of maximum to average angular rates. For simplicity, we use this
estimate to state complexity bounds for state transformations where
relevant, with the understanding that local-cost-sensitive estimates
can be obtained if needed.

Let $C_{{\rm max}}=\sup\{C(s_1,s_2)\;|\; a\leq s_1<s_2\leq b\}$,
$v_{{\rm max}}=\sup\{(L(s_2)-L(s_1))/(s_2-s_1)\;|\; a\leq s_1<s_2\leq b,
L(s_2)-L(s_1)>\theta\}$ and $v_{{\rm avg}} = (L(b)-L(a))/(b-a)$. 
If $L\leq\theta$, define $v_{{\rm max}} = v_{{\rm avg}}$.

\begin{lem}\label{lem:bit_cost_rough}
If $C(c,d)=0$ for $L(d)-L(c)\leq\theta$, then
\begin{equation}
C({\rm BIT}(a,b))\leq \frac{2(L(b)-L(a))}{\theta}
   \(\log_2\frac{v_{{\rm max}}}{v_{{\rm avg}}}+3\)C_{{\rm max}}\;.
\end{equation}
\end{lem}
The proof is given in Appendix~\ref{app:rtc}.

Let $v_{{\rm max}}$ and $v_{{\rm avg}}$ be as defined in
Lemma~\ref{lem:bit_cost_rough}, where by default $a=0$, $b=1$ and
$\theta$ is clear from context if not specified. Note that finiteness
of $v_{{\rm max}}$ requires that the path has no jumps of angular
distance $\theta$ or more.

\begin{thm}\label{thm:unknown_eigenphases}
Suppose that $\varphi_{s}$ is a $(1-4\gamma,\delta)$-dominant
eigenphase of $U_s$ in $\ket{\psi_r}$ for all $r<s$, where
$1-4\gamma\geq 2/3$. Let $\delta'=\min(\delta/2,\Delta/4)$,
$\theta<\arccos(\sqrt{1-\gamma})$ and
$C=2L(\log_2(v_{{\rm max}}/v_{{\rm avg}})+3)/\theta+1$. Then we can
transform $\ket{\psi_0}^{\tensor r}$ into $\ket{\psi_1}^{\tensor r}$
with an error amplitude bounded by
$6Ce^{-r\gamma^2/36}$ with at most $4rC$ instances of
phase estimation oracles with precision at least $\delta'$, an average
number of reflections bounded by $\langle n\rangle\leq 5rC$ and
$\langle\Gamma^n\rangle\leq\Gamma^{7rC}$ for $1\leq\Gamma\leq 7/4$.
\end{thm}

\begin{proof}
To implement the transformation, we apply $T_{px}$ of
Lemma~\ref{lem:tbr_no_phase_unconstrained} recursively to the
intervals of the BIT, terminating at intervals where the
transformation succeeds. The lemma guarantees that the transformation
succeeds if the angular length of the interval being tried is less
than $\arccos(\sqrt{1-\gamma})$. 

To determine the complexity of the transformation, we need to consider
a modified tree cost. Let $C_\theta(a,b) = C(a,(a+b)/2)+C((a+b)/2,b)$
if $L(b)-L(a)>\theta$ and $C_\theta(a,b)=0$ otherwise. Define
\begin{equation}
\tilde C_\theta({\rm BIT})(a,b) = C(a,b)+ C_\theta({\rm BIT})(a,b)\;.
\end{equation}
This accounts for the fact that the shortest intervals in the tree
that require action can be associated with arbitrarily short angular
lengths. It is their parents whose angular length must be too long for
terminating the recursion. With $C(a,b) = 1$, $\tilde C_\theta({\rm
  BIT})(0,1)$ is an upper bound on the number of intervals for which
transformation is attempted. According to
Lemma~\ref{lem:bit_cost_rough}, this is bounded by
$C=2L(\log_2(v_{{\rm max}}/v_{{\rm avg}})+3)/\theta+1$. The rest
follows by multiplying the complexities in
Lemma~\ref{lem:tbr_no_phase_unconstrained} by $C$ and applying
Lemma~\ref{lem:ldev1}. For the error amplitude we used amplitude
addition.
\end{proof}

The next theorem can be applied when little information on overlaps is
available, but we know sufficient eigenphase ranges for performing the
necessary reflections.  For this we need to consider the case where
the transformation has probability of success $p_s<1$ when $L(b)-L(a)
\leq\theta$. In this case, the process of subdividing $[a,b]$ may
continue indefinitely.  For $p_s>1/2$, the expected number of
intervals considered is finite with an exponentially decreasing tail
probability. This follows from the theory of Galton-Watson processes,
but in Appendix~\ref{app:gw} we give a statement and proof sufficient
for our purposes.

\begin{thm}\label{thm:unknown_overlaps}
Let $\theta=\arccos(\sqrt{1/3})\approx 0.96$.  Suppose that we know
phases $\tilde\varphi_{s}$ satisfying
$\tilde\varphi_{s}-\varphi_{s}\in[-\Delta/4,\Delta/4]$.  We can
transform $\ket{\psi_0}$ into $\ket{\psi_1}$ using $\langle n\rangle
\leq \bar n = 40\,L(\log_2(v_{{\rm max}}/v_{{\rm avg}})+3)/\theta+10$
reflections, with $\langle\Gamma^n\rangle\leq\Gamma^{36\bar n}$ for
$1\leq\Gamma\leq 14/13$.
\end{thm}

\begin{proof}
To implement the transformation, we apply $T'_{mx}$ to the intervals
of the BIT recursively. ($T'_{mx}$ is defined after Cor.~\ref{cor:txprime}.) The recursion terminates when a transformation
succeeds. For the intervals of angular length greater than $\theta$,
Cor.~\ref{cor:txprime} characterizes the distribution of the number of
reflections used whether or not the transformation succeeds. For
intervals $I$ of angular length at most $\theta$, the number of
subintervals that need to be tried is characterized by
Lemma~\ref{lem:galton_watson}, where the success probability satisfies
$p_s\geq 19/20$. Whether an interval needs to be tried depends only on
whether any of the intervals containing it (that is, above it in the BIT)
succeeded. Because of Cor.~\ref{cor:txprime_refinement},
Lemma~\ref{lem:ldev2} can be applied. Thus, according to
Cor.~\ref{cor:txprime}, the total number $n_I$ of reflections used in
our transforming across $I$ satisfies $\langle n_I\rangle <
9p_s/(2p_s-1)\leq 10$ and $\langle \Gamma^{n_I}\rangle
\leq\Gamma^{11(1+2/(2p_s-1))}\leq\Gamma^{36}$ for $1\leq\Gamma\leq
14/13 <
\min\{(1/(2\sqrt{p_s(1-p_s)}))^{1/11},8/7\}$. To finish the proof,
as was noted in the proof of Thm.~\ref{thm:unknown_eigenphases}, the
number of intervals of the BIT whose parents have angular length
greater than $\theta$ is bounded by $C=4L(\log_2(v_{{\rm max}}/v_{{\rm
avg}})+3)/\theta+1$. These intervals form a subtree ${\rm
BIT}_\theta$. Every reflection can be associated with the smallest
interval in ${\rm BIT}_\theta$ for whose traversal it was used. The statistics of the number of reflections associated
with each such node are bounded by the ones we obtained for intervals
of angular length at most $\theta$. Thus, we can apply
Lemma~\ref{lem:ldev1} to complete the proof.
\ignore{
\begin{verbatim}
% Octave:
ps = 19/20;
9*ps/(2*ps-1)
% = 9.5
(6+5)*(1+2/(2*ps-1))
% = 35.444 
function gamma = gw_gamma(ps)
  gamma = 1/(2*sqrt(ps*(1-ps))) 
endfunction
gw_gamma(19/20)^(1/(6+5))
% = 1.0784
14/13
% = 1/0769
\end{verbatim}
}
\end{proof}

\section{Summary of Complexities}
\label{sec:sd}

The most salient complexities are summarized in
Table~\ref{table:summary}. The bounds apply uniformly for $\bar
L=\max(\pi/2,L)$, $0<\Delta\leq\pi$, and $0<\epsilon<1/2$. We also
define $\cL = \bar L  \(\log\left( v_{{\rm max}}/ v_{{\rm avg}}\right)+2\)$. In order to
obtain the bounds, it is necessary to take into account the error
amplitudes contributed by two sources and make sure they do not exceed
the error goal of the algorithm. The first is in the implementation of
phase estimation and reflection oracles, and the second in our
multi-copy transformations. Calls to either oracle in our algorithms
require $\cO(\log(1/\delta)/\Delta)$ uses of the underlying unitary
operator for error amplitude $\delta$. If the state transformation
requires $M$ phase estimations or reflections, then we can set $\delta
= \epsilon/(2M)$ to ensure that the total error is bounded by
$\epsilon/2$, since error amplitudes are sub-additive. Thus the
complexity in terms of uses of the relevant unitary operators is
$\cO(M\log(M/\epsilon)/\Delta)$. The additional error in our
multi-copy state transfer algorithm is bounded by $e^{-\Omega(r)}$ per
state transformation attempt and is given explicitly in
Theorems~\ref{thm:simple_path_no_phase}
and~\ref{thm:unknown_eigenphases}. In our algorithms, the number
of phase estimation and reflection oracle calls per copy is linearly
related to the number $n$ of state transformation attempts. The latter
determines the total error from this contribution, which is
$ne^{-\Omega(r)}$. Thus, for an error goal of $\epsilon/2$ we can set
$r= \Theta(\log(n/\epsilon))$. This requires
$\cO(n\log(n/\epsilon))$ total phase estimation and reflection oracle
calls. After accounting for the error in the implementation of these
oracle calls, we obtain a complexity per copy of
$\cO(n\log(n\log(n/\epsilon)/\epsilon)/\Delta) =
\cO(n(\log(n/\epsilon)+\log\log(n/\epsilon))/\Delta) =
\cO(n\log(n/\epsilon)\Delta)$. 

\begin{table}
\begin{tabular}{|c|c|c|c||c||c|c|l|}
\hline
\multicolumn{4}{|c||}{Knowledge assumed}
      &\multicolumn{3}{c|}{}
      \\\hline
Overlap&
  Overlap&
    Eigenphase&
      Eigenphase&
Cost per copy& Number of copies& Reference\\
approx-& lower & ranges?& dominance?
  &&required&\\
imations?&bounds?&&&&&\\
\hline\hline
$\surd$ &&  $\surd$ & &
$\strutlike{\Bigg(}\displaystyle
  \cO\left(\frac{\bar L}{\Delta}\log\left(\frac{\bar L}{\epsilon}\right)\right)$ & 1 & 
  \begin{tabular}{c}
  Thm.~\ref{thm:simple_path},\\ Lemma~\ref{lem:n_overlap_checks}
  \end{tabular}
\\\hline
&$\surd$&&&
$\strutlike{\Bigg(}\displaystyle
 \cO\left(\frac{n}{\Delta}\log\left(\frac{n}{\epsilon}\right)\right)$
 & $\displaystyle
    \Theta\left(\log\left(\frac{n}{\epsilon}\right)\right)$ &
   Thm.~\ref{thm:simple_path_no_phase}
\\\hline
 &&  $\surd$ & & 
$ \strutlike{\Bigg(}\displaystyle
  \cO\left(\frac{\cL}{\Delta}\left(\log\left(\frac{\cL}{\epsilon}\right)\right)\right.
$
 & 1 & Thm.~\ref{thm:unknown_overlaps}
\\\hline
 &&  & $\surd$ &
$ \strutlike{\Bigg(}\displaystyle
  \cO\left(\frac{\cL}{\Delta}\left(\log\left(\frac{\cL}{\epsilon}\right)\right)\right.
$
 &
$\strutlike{\Bigg(}\displaystyle
  \Theta\left(\log\left(\frac{\cL}{\epsilon}\right)\right)$
 & Thm.~\ref{thm:unknown_eigenphases}
\\
\hline
\end{tabular}
\caption{Path transformation complexities. The entry in the column
``number of copies required'' gives the minimum needed by the
referred-to algorithm to achieve the desired error amplitude. In this
case, the error amplitude applies to all copies simultaneously, so
unless there are strong error correlations, individual copies may have
substantially less error. We have not determined the extent to which
the transformations of the copies can be parallelized. If eigenphase
dominance applies, we assume $\Delta$ is also a lower bound on the
resolution for dominance. The maximum angular velocity $v_{{\rm max}}$
can be bounded with $\theta=\Omega(1)$ in
Lemma~\ref{lem:bit_cost_rough}. The results in Appendix~\ref{app:are} may
also be helpful. $n$ is determined by $0=s_0<\ldots<s_n=1$ where
$p_{s_{l},s_{l+1}}>1/2+\Omega(1)$. We have $L=\cO(n)$, but $n$ could
be substantially larger than $L$ if we do not use preprocessing as in
Lemma~\ref{lem:n_overlap_checks}.}
\label{table:summary}
\end{table}

The formal meaning of the columns in Table~\ref{table:summary} for
assumed knowledge can be determined from the statements of the
referenced lemmas and theorems. Having knowledge of overlap
approximations means knowing enough about the $p_{s,t}$ to be able to
pick $0=s_0<\ldots<s_n=1$ such that $p_{s_{j-1},s_{j}}$ is large but
bounded away from $1$. This ensures that transforming along the $s_i$
is possible and efficient in terms of path length. Knowing overlap
lower bounds ensures the former only. When we say that eigenphase
ranges are known, we mean that for all $s$, we know an interval (or
more generally, a set) containing $\varphi_s$ such that the distance
from this interval to every other eigenphase of $U_s$ is 
at least $\Delta$. This is sufficient for implementing the
reflections with low error. The eigenphase dominance condition ensures
that we can statistically distinguish the wanted eigenphase when using
multiple copies of the states to infer adequate eigenphase ranges. The
formal definition for a path is in Thm.~\ref{thm:unknown_eigenphases}
based on Def.~\ref{dfn:dominant}.

Lemma~\ref{lem:n_overlap_checks} can be used to preprocess the
transformation steps so that the complexity of the first row of 
Table~\ref{table:summary} applies to subsequent transformations. Use of
Lemma~\ref{lem:n_overlap_checks} requires $\cO(n)$ calls to overlap
oracles, where $n$ is the number of actual transformation steps used.
Thus, the complexity has an additive term of $\cO(n\log(n/\epsilon)^2)$,
according to the note after Def.~\ref{dfn:overlap_oracle}. However, if
the recursive subdivision technique is used as in the last two rows of
the table, the use of overlap oracles can be avoided.

We have given the key complexities in terms of global quantities that
are simple to state. The complexities actually depend on local aspects
of the path. In particular, if the gaps $\Delta_s$ are typically large compared
to the minimum, sections of the path can be traversed much more quickly.
This can be taken into account by a finer complexity analysis, for example
by taking advantage of Lemma~\ref{lem:bit_cost}.

Our analyses apply to paths of non-degenerate eigenstates, but much of
it can be extended to paths of eigenspaces as follows. Suppose that
the path is characterized by a family of spaces $Z_s$
consisting of a union of eigenspaces of $U_s$ whose set of eigenphases have a
minimum distance $\Delta_s$ to all eigenphases of states orthogonal to
$Z_s$. The multi-copy transformation algorithms used when we
have insufficient information about the eigenphases require that $Z_s$
is an eigenspace. Reflections around $Z_s$ can be implemented with
the functional calculus of $Z_s$ as noted after
Def.~\ref{dfn:reflection_oracle}. The goal is to transform an
arbitrary initial state $\ket{\psi_0}\in Z_0$ into some $\ket{\psi_1}\in
Z_1$. To generalize our analysis, it is necessary to redefine the path
length. Let $\Pi_s$ be the projector onto $Z_s$. We define $L([a,b]) =
\sup L(a=s_0<\ldots<s_n=b)$, where
$L(s_0<\ldots<s_n)=\sum_{j=0}^{n-1}\theta(s_j,s_{j+1})$ and
$\theta(s,t) = \max\{\arccos(|\Pi_{t}{\psi_s}|)\,|\, \psi_s\in Z_s
\}$. Note that having no large jumps in the path implies that the
dimension of $Z_s$ is non-decreasing. The basic transformation steps
are the same, but their analysis requires the observation that the
reflections around the subspaces $Z_s$ and $Z_t$ 
are a direct sum of reflections on two-dimensional subspaces of the
space spanned by $Z_s$ and $Z_t$, see Ref.~\cite{szegedy_quantum_2004}.
Within each such subspace, the transformation behaves as expected. The
relevant overlaps now depend on the relationship between the
reflection axes in the mentioned two-dimensional subspaces.

\acknowledgments We thank A. Harrow for discussions regarding the
algorithm in the unknown-eigenphase case. This research was supported
by the Perimeter Institute for Theoretical Physics, by the Government of
Canada through Industry Canada and by the Province of Ontario through
the Ministry of Research and Innovation. Contributions to this work by
NIST, an agency of the US government, are not subject to copyright
laws. This work was supported by the National Science Foundation under
grant PHY-0803371 through the Institute for Quantum Information at the
California Institute of Technology. We also thank the Laboratory
Directed Research and Development Program at Los Alamos National
Laboratory for support.

\appendix

\section{Proof of Lemma~\ref{lem:ldev2}}
\label{app:pldev2}
Let $\mu$ be the measure for the probability distribution of its
arguments. The conditional independence assumption is equivalent to
having the $C_j$ and $W_j$ generated via the sequence of probabilistic
transitions $\ldots \rightarrow (C_j,\mathbf{W}_j) \rightarrow
(\mathbf{W}_j) \rightarrow (C_{j+1},\mathbf{W}_{j+1})
\rightarrow\ldots$. It follows that $C_j$ is conditionally independent
of the $C_k$ for $k < j$ given $\mathbf{W}_{j}$. Formally we 
have
\begin{eqnarray*}
d\mu(\mathbf{C}_k,\mathbf{W}_k) &=&
  d\mu(C_k,W_k,\mathbf{C}_{k-1}|\mathbf{W}_{k-1})d\mu(\mathbf{W}_{k-1}) \\
  &=&
  d\mu(C_k,W_k|\mathbf{W}_{k-1})d\mu(\mathbf{C}_{k-1}|\mathbf{W}_{k-1})
    d\mu(\mathbf{W}_{k-1})\\
  &=&
  d\mu(C_k,W_k|\mathbf{W}_{k-1})d\mu(\mathbf{C}_{k-1},\mathbf{W}_{k-1})\\
  &=&
  d\mu(C_k|\mathbf{W}_{k})d\mu(W_k|\mathbf{W}_{k-1})d\mu(\mathbf{C}_{k-1},\mathbf{W}_{k-1})\\
  &\vdots&\\
  &=&
  \left(\prod_{j=1}^k d\mu(C_j|\mathbf{W}_{j})d\mu(W_j|\mathbf{W}_{j-1})\right)\\
  &=& \left(\prod_{j=1}^k d\mu(C_j|\mathbf{W}_{j})\right)d\mu(\mathbf{W}_k)\;,
\end{eqnarray*}
where the omitted identities involve applying the first steps
recursively to the last term.

Given the constraints on $\Gamma$, we obtain
\begin{eqnarray*}
\langle\Gamma^{C_{{\rm tot},k}}\rangle &=& 
  \int \Gamma^{\sum_{j=1}^k C_j}d\mu(\mathbf{C_k},\mathbf{W_k}) \\ 
  &=& 
  \int \left(\prod_{j=1}^k \int\Gamma^{C_j}d\mu(C_j|\mathbf{W_j})\right)d\mu(\mathbf{W_k}) \\
  &=&
  \int \left(\prod_{j=1}^k \int\left(1-V_j + V_j\int\Gamma^{C_j}d\mu(C_j|\mathbf{W_j},V_j)\right)d\mu(V_j|\mathbf{W_j})\right)d\mu(\mathbf{W_k}) \\
  &\leq&
  \int\left(\prod_{j=1}^k \int (1-V_j + V_j\Gamma^{\tilde C})d\mu(V_j|\mathbf{W_j})\right)d\mu(\mathbf{W_k})\\
  &=&
  \int\left(\prod_{j=1}^k \int \Gamma^{\tilde CV_j}d\mu(V_j|\mathbf{W_j})\right)d\mu(\mathbf{W_k})\\
  &=&
  \int\left(\prod_{j=1}^k \int \Gamma^{\tilde CV_j}d\mu(C_j|\mathbf{W_j})\right)d\mu(\mathbf{W_k})\\
  &=&
  \int \Gamma^{\tilde C\sum_{j=1}^k V_j}d\mu(\mathbf{C}_k,\mathbf{W}_k)\\
  &=& \int\Gamma^{\tilde C\sum_{j=1}^k V_j}d\mu(\mathbf{C}_k)\;,
\end{eqnarray*}
because $V_j$ is a function of $C_j$. To get the desired bound, we let
$k\rightarrow \infty$, use the monotone convergence theorem, and apply
the bound on $\langle \Lambda^{m}\rangle$ with $\Lambda=\Gamma^{\tilde
  C}$:
\begin{eqnarray*}
  \lim_{k \rightarrow \infty} \int\Gamma^{\tilde C\sum_{j=1}^k V_j}d\mu(\mathbf{C}_k)   &=&  \int \lim_{k \rightarrow \infty}  \Gamma^{\tilde C\sum_{j=1}^k V_j}d\mu(\mathbf{C}_k) \\
  &=& \langle \Gamma^{\tilde C m} \rangle \le \Lambda ^{\tilde m}\;.
\end{eqnarray*}

\section{Recursive transformations complexity}
\label{app:rtc}

Let $L(s)$ be the length of the path $\ket{\psi_t}$ from $t=0$ to
$t=s$. Define $\underline s(l)=\inf \{s:L(s)\geq l,s\in[0,1]\}$ and
$\overline s(l)=\sup \{s:L(s)\leq l,s\in[0,1]\}$. We have $\underline
s(l)\leq \overline s(l)$ and the state is constant on the open
interval $(\underline s(l),\overline s(l))$. The functions $\underline s$ and
$\overline s$ are monotone. Given $l\in [L(a),L(b)]$ and a distance scale
$\theta$, we define a local maximum speed variation at $l$ by
\begin{equation}\label{eq:s_theta}
  \sigma_\theta(l,[a,b]) =
    \begin{array}[t]{l@{}l@{}l}\displaystyle
    \sup&\Big\{\frac{s_4-s_1}{s_3-s_2}\;|\;&
          a\leq s_1\leq s_2<s_3\leq s_4\leq b,\\
          && L(s_2)<l<L(s_3),\\
          && L(s_3)-L(s_2)>\theta,\\
          &&L(s_4)-L(s_1)\leq 2\theta\Big\}\;.
    \end{array}
\end{equation}
If the set under the supremum is empty, let $\sigma_{\theta}(l,[a,b])=1$.
To justify the description of $\sigma_\theta$ as a speed variation, we
define average speeds between $l_1$ and $l_2>l_1$ by $\overline
v(l_1,l_2) = (l_2-l_1)/(\underline s(l_2)-\overline s(l_1))$ and
$\underline v(l_1,l_2) = (l_2-l_1)/(\overline s(l_2)-\underline
s(l_1))$, allowing for constant sections on the path. A more direct
local speed variation is 
\begin{equation}
\rho_\theta(l,[a,b]) = 
  \begin{array}[t]{l@{}l@{}l}\displaystyle
  \sup  \Big\{&\frac{\overline v(l_2,l_3)}{\underline v(l_1,l_4)}\;|\;& 
              L(a)\leq l_1\leq l_2<l<l_3\leq l_4,\\
              &&l_3=l_2+\theta,\\
              &&l_4=\min(l_1+2\theta,L(b))
              \Big\}\;.
  \end{array}
\end{equation}
If the set under the supremum is empty, let
$\rho_{\theta}(l,[a,b])=1$. The description of $\sigma_\theta(l,[a,b])$ as
a speed variation at $l$ at scale $\theta$ comes from the observation
that
\begin{lem}For all intervals $I$
\begin{equation}
\sigma_\theta(l,I)\leq 2\rho_\theta(l,I)
\label{eq:sigma_leq_2rho}
\end{equation}
with equality if $l\in[L(a)+2\theta,L(b)-2\theta]$, the path
is continuous and has no constant intervals.
\end{lem}

\begin{proof}
The set $X$ in the definition of $\sigma_\theta(l,I)$ is empty iff
$L(b)-L(a)\leq\theta$ or $l\notin [L(a),L(b)]$. If the latter holds,
then the set $Y$ in the definition of $\rho_\theta(l,I)$ is empty.
If not, then either $Y$ is empty or $L(b)=L(a)+\theta$. In this
case $Y = \{y\}$ with $y\geq 1$. Consider $r=(s_4-s_1)/(s_3-s_2)\in X$,
with $s_i$ as in Eq.~\ref{eq:s_theta}. Let $l_i=L(s_i)$
and define $l_1'=l_1$, $l_4'=\min(l_1+2\theta,L(b))\geq l_4$
and choose $l_2'$, $l_3'$ such that $l_3'=l_2'+\theta$,
$l_2\leq l_2'<l<l_3'\leq l_3$. Then
\begin{eqnarray*}
r &\leq& \frac{\overline s(l_4')-\underline s(l_1')}{
               \underline s(l_3')-\overline s(l_2')}\\
  &\leq& 2\left(\frac{l_3'-l_2'}{l_4'-l_1'} \right)
           \frac{\overline s(l_4')-\underline s(l_1')}{
               \underline s(l_3')-\overline s(l_2')}\\
  &\leq& 2 \rho_\theta(l,I).
\end{eqnarray*}
For the reverse inequality, given $\epsilon>0$ arbitrarily small,
choose $l_i$ such that
$r'=\overline{v}(l_2,l_3)/\underline{v}(l_1,l_4) \geq
\rho_\theta(l,I)-\epsilon$. The constraint on $l$ implies that
$l_4=l_1+2\theta$, so $r'=2(\overline s(l_4)-\underline
s(l_1))/(\underline s(l_3)-\overline s(l_2))$. Let $\delta>0$ be
arbitrarily small. Let $s_i=\overline s(l_i)$ for $i=2,4$ and
$s_j=\underline s(l_j)$ for $j=1,3$. Continuity implies that
$L(s_i)=l_i$. If $l_2>l_1$, then $s_2>s_1$ and we let
$s'_2=s_2-\delta$. Else $l_3<l_4$ and we let $s'_3=s_3+\delta$.
Unassigned $s'_j$ are set to $s_j$. The assumptions imply that for
$\delta$ small enough, the $s'_i$ satisfy the constraints in
Eq.~\ref{eq:s_theta}, showing that $\sigma_\theta(l,I)\geq
2\rho_\theta(l,I)-\epsilon$. Letting $\epsilon\downarrow 0$ gives the
desired result.
\end{proof}

The next lemma gives
a bound on the cost of a symmetric binary interval tree that is sensitive to local variations.
\begin{lem}\label{lem:bit_cost}
If $C(c,d)=0$ for $L(d)-L(c)\leq\theta$,
then 
\begin{eqnarray}
C({\rm BIT}(a,b)) &\leq& \frac{1}{\theta}\sum_{k=0}^\infty
  \frac{1}{2^k} \int_{L(a)}^{L(b)} 
  \Big(\big\lfloor\log_2\big(\sigma_{2^k\theta}(l,[a,b])\big)\big\rfloor+1\Big)\overline C_{2^k\theta}(l)
  dl\;,\nonumber\\
\end{eqnarray}
where $\overline C_{\theta'}(l) = \sup\{ C(s_1,s_2)\;|\; 
L(s_1)<l<L(s_2), \theta'<L(s_2)-L(s_1)\leq 2\theta'\}$.
\end{lem}
\begin{proof}
The bound is obtained in three steps. First we separately sum over
intervals $[s_1,s_2]$ in the tree in each length class
$2^k\theta<L(s_2)-L(s_1)\leq 2^{k+1}\theta$. Second, we uniformly,
randomly assign the cost $C(s_1,s_2)$ to the open interval between
$L(s_1)$ and $L(s_2)$ and integrate over the length variable.
Third, we use bounds on costs and numbers of intervals in the tree
spanning the value of the length variable. The steps are implemented
in the following sequence of identities and inequalities:
\begin{eqnarray*}
C({\rm BIT}(a,b)) 
  &=& \sum_{[s_1,s_2]\in{\rm BIT}(a,b)} C(s_1,s_2)\\
  &=& \sum_{k=0}^{\infty}
        \sum{}_{\begin{array}{l}\scriptstyle
              [s_1,s_2]\in{\rm BIT}(a,b), \\\scriptstyle
              2^k\theta<L(s_2)-L(s_1)\leq 2^{k+1}\theta\end{array}}
           C(s_1,s_2)\\
  &=&\sum_{k=0}^{\infty}
        \sum{}_{\begin{array}{l}\scriptstyle
              [s_1,s_2]\in{\rm BIT}(a,b), \\\scriptstyle
              2^k\theta<L(s_2)-L(s_1)\leq 2^{k+1}\theta\end{array}}
           C(s_1,s_2)\frac{1}{L(s_2)-L(s_1)}\int_{L(s_1)}^{L(s_2)}dl\\
  &\le&\sum_{k=0}^{\infty}
        \sum{}_{\begin{array}{l}\scriptstyle
              [s_1,s_2]\in{\rm BIT}(a,b), \\\scriptstyle
              2^k\theta<L(s_2)-L(s_1)\leq 2^{k+1}\theta\end{array}}
           C(s_1,s_2)\frac{1}{2^k \theta}\int_{L(a)}^{L(b)}\mathbf{1}_{[L(s_1),L(s_2)]}(l)dl\\
  &\le&\sum_{k=0}^{\infty}
       \frac{1}{2^k \theta}\int_{L(a)}^{L(b)}
      \Bigg( \sum{}_{\begin{array}{l}
                  \scriptstyle
                  [s_1,s_2]\in{\rm BIT}(a,b),\\
                  \scriptstyle
                  2^k\theta<L(s_2)-L(s_1)\leq 2^{k+1}\theta\end{array}}
          \mathbf{1}_{[L(s_1),L(s_2)]}(l)\Bigg)\overline C_{2^k\theta}(l)dl
\end{eqnarray*}
To finish the proof, note that the number in parenthesis in the last line,
\begin{eqnarray*}
  \bigg| \Big\{[s_1,s_2]\in{\rm BIT}(a,b)\;|\;
                   L(s_1)<l<L(s_2), 
                   2^k\theta< L(s_2)-L(s_1)\leq 2^{k+1}\theta\Big\}
              \bigg|\;,
\end{eqnarray*}
is the size of a set 
of nested intervals in the tree. Let $[s_1,s_4]$ be the
biggest and $[s_2,s_3]$ the smallest of these intervals. Because of
the way the tree is constructed, the number of these intervals is
given by
\begin{eqnarray*}
  \log_2((s_4-s_1)/(s_3-s_2))+1 \le \sigma_{2^k\theta}+1\;,
\end{eqnarray*}
where the inequality follows from the definition of $ \sigma_{2^k\theta}$\;.
\end{proof}

Finally, we give the proof of Lemma~\ref{lem:bit_cost_rough} stated in
the text.

\begin{proof}[Proof of Lemma~\ref{lem:bit_cost_rough}]
We begin by bounding
$\int_{L(a)}^{L(b)}\log_2(\sigma_{\theta'}(l,[a,b]))dl$ for $\theta'\geq
\theta$ in terms of the maximum and average angular rates. If
$L(b)-L(a)\leq\theta'$, then $\sigma_{\theta'}(l,[a,b])=1$ so the integral
is $0$. Assume $L(b)-L(a)>\theta'$ and let $\epsilon>0$ be
arbitrarily small. For every $l$, choose $a\leq s_1(l)\leq
s_2(l)<s_3(l)\leq s_4(l)\leq b$ such that the following hold: The
$s_i=s_i(l)$ satisfy the constraints given in the 
definition of $\sigma_{\theta'}(l,[a,b])$ (see Eq.~\ref{eq:s_theta}),
$\sigma_{\theta'}(l,[a,b])\leq 2^{\epsilon}(s_4(l)-s_1(l))/(s_3(l)-s_1(l))$,
and the $s_i(l)$ are measurable functions. We can now bound
\begin{eqnarray*}
\int_{L(a)}^{L(b)}\log_2(\sigma_{\theta'}(l,[a,b]))dl
  &\leq& (L(b)-L(a))\epsilon 
         + \int_{L(a)}^{L(b)}\log_2 \frac{s_4(l)-s_1(l)}{s_3(l)-s_2(l)} dl\;.
\end{eqnarray*}
Jensen's inequality, applied to the concave $\log_2$ function, gives
\begin{eqnarray*}
\int_{L(a)}^{L(b)}\log_2 \frac{s_4(l)-s_1(l)}{s_3(l)-s_2(l)} dl
  &\leq& (L(b)-L(a))\log_2 \frac{1}{L(b)-L(a)}\int_{L(a)}^{L(b)}
 \frac{s_4(l)-s_1(l)}{s_3(l)-s_2(l)} dl \;.
\end{eqnarray*}
We bound the denominator inside the integral with the inequality
$(s_3(l)-s_2(l))>\theta'/v_{{\rm max}}$, which does not depend on the
variable of integration, so we can continue by bounding the integral
of the numerator.  We first change the order of integration
\begin{eqnarray*} \int_{L(a)}^{L(b)}\( s_4(l)-s_1(l)\) dl
  &=& \int_{L(a)}^{L(b)} dl\int_{s_1(l)}^{s_4(l)} dt
  = \int_{a}^{b}dt \int_{L(a)}^{L(b)} \mathbf{1}_{\{l| t\in[s_1(l),s_4(l)]\}}(l)dl\;.
\end{eqnarray*}
If $t\in[s_1(l),s_4(l)]$, then $L(l)\leq L(s_4(l))\leq
L(s_1(l))+2\theta'\leq L(t)+2\theta'$.  Similarly, $L(l)\geq
L(s_1(l))\geq L(s_4(l))-2\theta'\geq L(t)-2\theta'$.  It follows that
the inner integral is bounded by $4\theta'$.  This gives the bound
\begin{eqnarray*} \frac{1}{L(b)-L(a)}\int_{L(a)}^{L(b)}( s_4(l)-s_1(l)) dl
   &\leq& \frac{1}{L(b)-L(a)}\int_{a}^{b}4\theta' dt
  = \frac{4\theta'(b-a)}{L(b)-L(a)}
   = \frac{4\theta'}{v_{{\rm avg}}}\;.
\end{eqnarray*}

Combining the bounds and letting $\epsilon$ go to $0$, we obtain
\begin{equation*}
\int_{L(a)}^{L(b)}\log_2(\sigma_{\theta'}(l,[a,b]))
  \leq (L(b)-L(a))\(\log_2\frac{v_{{\rm max}}}{v_{{\rm avg}}}+2\)\;.
\end{equation*}
Substituting into the bound of Lemma~\ref{lem:bit_cost}, we get
\begin{eqnarray*}
C({\rm BIT}(a,b))&\leq &
\frac{1}{\theta}\sum_{k=0}^{\infty} \frac{1}{2^k} 
     (L(b)-L(a))\(\log_2\frac{v_{{\rm max}}}{v_{{\rm avg}}}+3\)C_{{\rm max}}\\
  &\leq&
    \frac{2(L(b)-L(a))}{\theta}\(\log_2\frac{v_{{\rm max}}}{v_{{\rm avg}}}+3\)C_{{\rm max}}\;,
\end{eqnarray*}
as desired.
\end{proof}

\section{A Galton-Watson lemma}
\label{app:gw}

\begin{lem}\label{lem:galton_watson}
Consider the random process starting from $S =
\{([a,b],{\rm active})\}$ which is defined recursively as follows: For
all $N=([c,d],{\rm active})$ in $S$, replace $N$ with with
$([c,d],{\rm inactive})$ and with probability $1-p_s$ add
$([c,(c+d)/2],{\rm active})$ and $([(c+d)/2,d],{\rm active})$ to $S$.
Let $S_\infty$ be the possibly infinite set obtained by running this
process countably many times. For $p_s>1/2$, the expected size of
$S_\infty$ is $\langle |S_\infty|\rangle = p_s/(2p_s-1)$
and $\langle \Gamma^{|S_\infty|}\rangle \leq  \Gamma^{1+2/(2p_s-1)}$ for
$1\leq\Gamma\leq 1/(2\sqrt{p_s(1-p_s)})$.
\end{lem}

\begin{proof} We outline the proof, omitting some necessary existence
arguments. Either $S_\infty = \{([a,b],{\rm inactive})\}$, which
happens with probability $p_s$, or $S_\infty =
\{([a,b],{\rm inactive})\}\cup S_l\cup S_r$, where $S_l$ and $S_r$ are
independent with the same statistics as $S_\infty$. The latter
happens with probability $1-p_s$. Thus $\langle |S_\infty|\rangle =
p_s + 2 (1-p_s) \langle |S_\infty|\rangle$, or $\langle
|S_\infty|\rangle = p_s/(2p_s-1)$. Similarly, using the independence
of $S_l$ and $S_r$ again, we find $\langle\Gamma^{|S_\infty|}\rangle =
\Gamma\left(p_s + (1-p_s)\bar
\langle\Gamma^{|S_\infty|}\rangle^2\right)$. This is solved by
\begin{equation*}
\langle\Gamma^{|S_\infty|}\rangle = \frac{1}{2(1-p_s)\Gamma}
   \left(
      1\pm\sqrt{1-4\Gamma^2p_s(1-p_s)}
   \right)\;.
\end{equation*}
The relevant solution is the negative branch, as can be seen by
checking that $\langle 1^{|S_\infty|}\rangle = 1$. The maximum value of
$\Gamma$ for which it is defined is $\Gamma=1/(2\sqrt{p_s(1-p_s)})$.
For this $\Gamma$, $\langle\Gamma^{|S_\infty|}\rangle=2p_s\Gamma$. The following sequence of
inequalities together with log-convexity completes the proof:
\begin{eqnarray*}
2p_s &=& 1+(2p_s-1)\\
     &\leq& 1+\frac{2p_s-1}{4p_s(1-p_s)}\\
     &=& 1+\( \frac{1}{2p_s-1}\)\(\frac{(2p_s-1)^2}{4p_s(1-p_s)}\)\\
     &\leq& \(1+\frac{(2p_s-1)^2}{4p_s(1-p_s)}\)^{\frac{1}{2p_s-1}}\\
     &=& \(\frac 1{4p_s(1-p_s)}\)^{\frac{1}{2p_s-1}} = \Gamma^{\frac 2{2p_s-1}}
\end{eqnarray*}
at the maximum $\Gamma$. 
\end{proof}

\section{Angular rate estimation}
\label{app:are}

Here are some tools for estimating average angular rates for paths of
eigenstates of normal operators.

\begin{lem}\label{lem:general_angle_bound}
Let $H$ be a normal operator with eigenstate $\ket{\psi}$ having
eigenvalue $\lambda$. Suppose $H+S$ is normal and has an eigenspace $V$
with eigenvalue $\lambda+\delta$ gapped by $\Delta$.
Then the maximum angle from $\ket{\psi}$ to $V$ is bounded by
$\arcsin(\|(S-\delta)\ket{\psi}\|/\Delta)$.
\end{lem}

\begin{proof}
Let $\Pi$ be the projector onto the orthogonal complement of $V$. It
suffices to show that $|\bra{\psi'}\Pi\ket{\psi}|\leq
\|(S-\delta)\ket{\psi}\|/\Delta$ for all $\ket{\psi'}$. We can use the
``resolvent trick'' to express
\begin{eqnarray*}
\Pi &=& \frac{1}{2\pi i}\int \frac{1}{z}-(z-((H+S)-(\lambda+\delta)))^{-1} dz\\
    &=& \frac{1}{2\pi i}\int (z-((H+S)-(\lambda+\delta)))^{-1} ((H+S)-(\lambda+\delta)) dz/z\;,
\end{eqnarray*}
where the integral is over a circle of radius $d$ less than $\Delta$
around $0$. Thus
\begin{eqnarray*}
|\bra{\psi'}\Pi\ket{\psi}| &=&
   \left|\bra{\psi'}\frac{1}{2\pi}\int  (z-((H+S)-(\lambda+\delta)))^{-1}(S-\delta) dz/z\ket{\psi}\right|\\
   &\leq&
      \|(S-\delta)\ket{\psi}\| (\Delta-d)^{-1}\;.
\end{eqnarray*}
The result follows by letting $d$ go to $0$.
\end{proof}

\begin{cor}
Let $H(s)$ be a family of normal operators,
$H(s)\ket{\psi(s)}=\lambda(s)\ket{\psi(s)}$ with all objects
differentiable at $s=t$. Let $\Pi_t^{\perp} =
\one-\ketbras{\psi(t)}{\psi(t)}{}$. If the eigenvalues $\lambda(s)$ are
gapped with gap $\Delta$ in a neighborhood of $s=t$, then
\begin{equation}
\left\|\Pi_t^{\perp}\frac{d\ket{\psi(s)}}{ds}\Bigg|_{s=t}\right\|
  \leq \left\|\frac{d H(s)}{ds}\Bigg|_{s=t}\right\|/\Delta\;.
\end{equation}
\end{cor}

\begin{proof}
It suffices to apply first-order perturbation theory to
Lemma~\ref{lem:general_angle_bound}, noting that $H(s) = H(t)+(t-s)(d
H(s)/ds|_{s=t}) + o(|t-s|)$ and $\lambda(s) = \lambda(t)+
(t-s)\bra{\psi(t)}(d H(s)/ds|_{s=t})\ket{\psi(t)} + o(|t-s|)$.
\end{proof}

\bibliographystyle{plain}
\bibliography{mbqc}
\end{document}